\documentclass[%
 aip,
 amsmath,amssymb,
 reprint,%
]{revtex4-1}

\usepackage{graphicx}
\usepackage{dcolumn}
\usepackage{bm}
\usepackage{xcolor}

\usepackage[utf8]{inputenc}
\usepackage[T1]{fontenc}
\usepackage{mathptmx}
\usepackage{etoolbox}
\usepackage{CJKutf8}

\makeatletter
\def\@email#1#2{%
 \endgroup
 \patchcmd{\titleblock@produce}
  {\frontmatter@RRAPformat}
  {\frontmatter@RRAPformat{\produce@RRAP{*#1\href{mailto:#2}{#2}}}\frontmatter@RRAPformat}
  {}{}
}%
\makeatother
\begin{document}

\preprint{AIP/123-QED}

\title{Dynamics of flexible fibers in \textcolor{black}{confined shear} flows at finite Reynolds numbers}

\author{Jian Su \textcolor{black}{(\begin{CJK*}{UTF8}{gbsn}苏建\end{CJK*})} }\thanks { These authors contributed equally.}
\affiliation{Technion – Israel Institute of Technology, Haifa, Israel, 3200003}
\affiliation{Department of Physics, Guangdong Technion - Israel Institute of Technology, 241 Daxue Road, Shantou, Guangdong, China, 515063}

\author{Kun Ma \textcolor{black}{(\begin{CJK*}{UTF8}{gbsn}马坤\end{CJK*})}} \thanks { These authors contributed equally.}
\affiliation{School of Mathematics, Sichuan University, Chengdu, Sichuan, China, 610065}%

\author{Zhongyu Yan \textcolor{black}{(\begin{CJK*}{UTF8}{gbsn}严钟渝\end{CJK*})}}
\affiliation{Department of Physics, Guangdong Technion - Israel Institute of Technology, 241 Daxue Road, Shantou, Guangdong, China, 515063}
\affiliation{Technion – Israel Institute of Technology, Haifa, Israel, 3200003}

\author{Qiaolin He \textcolor{black}{(\begin{CJK*}{UTF8}{gbsn}贺巧琳\end{CJK*})}} \thanks{Corresponding author: qlhejenny@scu.edu.cn}
\affiliation{School of Mathematics, Sichuan University, Chengdu, Sichuan, China, 610065}

\author{Xinpeng Xu \textcolor{black}{(\begin{CJK*}{UTF8}{gbsn}徐新鹏\end{CJK*})}}\thanks{Corresponding author: xu.xinpeng@gtiit.edu.cn}
\affiliation{Department of Physics, Guangdong Technion - Israel Institute of Technology, 241 Daxue Road, Shantou, Guangdong, China, 515063}
\affiliation{Technion – Israel Institute of Technology, Haifa, Israel, 3200003}


\date{\today}

\begin{abstract}
We carry out a numerical study on the dynamics of a single non-Brownian flexible fiber in two-dimensional \textcolor{black}{confined simple shear (Couette)} flows at finite Reynolds numbers. We employ the bead-spring model of flexible fibers to extend the fluid particle dynamics (FPD) method  that is originally developed for rigid particles in viscous \textcolor{black}{fluids}. We implement the extended FPD method using a multiple-relaxation-time (MRT) scheme of the lattice Boltzmann method (LBM). The numerical scheme is validated firstly by a series of benchmark simulations that involve \textcolor{black}{fluid}-solid coupling. The method is then used to study the dynamics of flexible fibers in Couette flows. We only consider the highly symmetric case where the fibers are placed on the symmetry center of Couette flows and we focus on the effects of the fiber stiffness, the confinement strength, and the finite Reynolds number (from 1 to 10). A diagram of the fiber shape is obtained. For fibers under weak confinement and a small Reynolds number, three distinct tumbling orbits have been identified. (1) Jeffery orbits of rigid fibers. The fibers behave like rigid rods and tumble periodically without any visible deformation. (2) S-turn orbits of slightly flexible fibers. The fiber is bent to an S-shape and is straightened again \textcolor{black} {when it orients to an angle of around 45 degrees relative to the positive x direction}. (3) S-coiled orbits of fairly flexible fibers. The fiber is folded to an S-shape and tumbles periodically and steadily without being straightened anymore during its rotation. Moreover, the fiber tumbling is found to be hindered by increasing either the Reynolds number or the confinement strength, or both. 
\end{abstract}

\maketitle


\section{Introduction}\label{Sec:introduction}

Flexible fibers immersed in viscous fluids are ubiquitous in many industrial and biological processes~\cite{du2019dynamics,duprat2015fluid,Du2019PRF}. In industry, fibers are the functional microstructure of many industrial complex fluids. The dynamics of suspended fibers is a major concern in many important processes such as lubrication, extrusion, and molding~\cite{hamedi2021}. In biology, microorganisms use fibrous flagella to swim and stir~\cite{lauga2009hydrodynamics}; eukaryotic cells control fibrous cytoskeletal networks to position their nuclei properly during their physiological processes~\cite{shelley2016dynamics}. 
It has already been known that the shape, orientation, and distribution of the fibers are the most important structural features that determine the collective behaviors and/or the rheological properties of the fiber suspensions. It is, therefore, important to understand the transport dynamics of fibers in viscous flows for the design and control of fiber suspension processing. 

The dynamics of rigid fibers in unconfined viscous flows at low Reynolds numbers have been extensively studied. The earliest theoretical studies date back to the pioneering works by G.B. Jeffery in 1922~\cite{jeffery1922motion} and by F.P. Bretherton~\cite{bretherton1962motion} in 1962, where rigid fibers of various aspect ratios in unconfined simple shear flows are shown to tumble or rotate following Jeffery orbits. These tumbling dynamics are recently found to be modified by the presence of confining walls~\cite{Du2019PRF,du2019dynamics}. In contrast, the transport dynamics of flexible fibers are surprisingly complex and exhibit very rich dynamics~\cite{du2019dynamics}, resulting from a coupling among rotation, deformation, and two-phase (\textcolor{black}{fluid}-solid) hydrodynamics. 
Recently, numerous theoretical and experimental studies have been done to investigate the dynamics, particularly the buckling instabilities~\cite{becker2001instability,kanchan2019numerical,slowicka2022buckling}, of flexible fibers in different ambient fluid flows, such as shear flows~\cite{forgacs1959particle,tang2005dynamic,khare2006cross,slowicka2015dynamics,kuei2015dynamics,liu2018morphological}, oscillatory shear flows~\cite{bonacci2023dynamics}, extensional flows~\cite{kantsler2012fluctuations}, cellular flows~\cite{young2007stretch}, and swirling flows~\cite{guo2009novel}. 
The fiber dynamics are known to be sensitive not only to the fiber bending stiffness but also to many other different factors, such as the strength of confinement, Reynolds number, shear rate, flow curvature, thermal fluctuations, and the initial conditions that lead to various modes of three-dimensional (3D) motion~\cite{skjetne1997simulation,slowicka2020flexible,slowicka2022buckling}. However, the effects of these factors on the transport dynamics of flexible fibers are rarely explored~\cite{Du2019PRF,du2019dynamics} and they pose great challenges to theoretical modeling and simulations. The major reasons include that (1) flexible fibers have many degrees of freedom in deformation and can exhibit microscopic instabilities, and (2) the suspension dynamics involve moving fluid-fiber interfaces and long-range hydrodynamic interactions. 




In the past decades, many numerical approaches have been proposed to simulate flexible fibers suspended in different viscous flows. Yamamoto and Matsuoka~\cite{yamamoto1993method} proposed a simple bead-spring chain model of flexible fibers to study the fiber dynamics in unconfined simple shear flows at low Reynolds numbers. Schmid \emph{et al.}~\cite{schmid2000mechanical} developed a similar model by regarding the fibers as a chain of rods connected by hinges, in which the rods, from the same or from different fibers, interact with each other through constraint, frictional, and lubrication forces. This model has been used to study the single-fiber dynamics in various flows, the collective fiber dynamics, and the rheological properties of fiber suspensions. In both types of fiber models, the only hydrodynamic forces considered are the viscous drag forces applied to the fibers from the prescribed background flows at low Reynolds numbers, in which the backward effects of fiber motion on the background flows and the long-range hydrodynamic interactions between fiber segments from the same or from different fibers have been neglected completely. Therefore, these models are applicable only to infinitely dilute fiber suspensions at low Reynolds numbers, where the effects of hydrodynamic interactions and inertia are both absent. 

To treat the many-body long-range hydrodynamic interactions that are important in single fiber dynamics under confinement or generally in more dense suspensions, \textcolor{black}{a number of numerical methods have been developed such as Immersed Boundary Method~\cite{feng2004immersed,kanchan2019numerical}, Fictitious Domain Method (FDM)~\cite{he2018least}, Dissipative Particle Dynamics (DPD)~\cite{lobaskin2004electrophoretic}, Stochastic Rotational Dynamics (SRD)~\cite{malevanets1999mesoscopic}, Stokesian Dynamics (SD)~\cite{brady1988stokesian}, Smoothed Profile Method (SPM)~\cite{yamamoto2001simulating}, Multiparticle Collision Dynamics (MPCD)~\cite{kapral2008multiparticle}, Hybrid Phase Field method for Fluid-Structure Interactions (HPFM)~\cite{hong2021hybrid}, and Fluid Particle Dynamics (FPD)~\cite{tanaka2000simulation,tanaka2006viscoelastic,tanaka2018physical}.} 
In addition, the effects of inertia at finite Reynolds numbers are present in many industrial applications and are known to dramatically affect flow behavior even at the single particle level~\cite{yan2007hydrodynamic,subramanian2006trajectory}. Zettner and Yoda~\cite{zettner2001moderate} studied experimentally the circular particles in Couette flows (\emph{i.e.}, confined simple shear flows) at finite particle-under-shear Reynolds number ${\rm Re_{ps}}$. Ding and Aidun~\cite{ding2000dynamics} employed the Lattice Boltzmann Method (LBM) to study the elliptical particles in simple shear flows and found that by
increasing ${\rm Re_{ps}}$, the tumbling period of the elliptical particle increases, and eventually becomes infinitely large at a critical Reynolds number, ${\rm {Re}}_{\rm {cr}}$. That is, for ${{\rm Re_{ps}}}>{\rm {Re}}_{\rm {cr}}$, the particle is stuck and becomes stationary in a steady-state flow. Near  ${\rm {Re}}_{\rm {cr}}$, the rotation period $T$ follows a universal scaling law: $\dot {\gamma} T \propto ({\rm {Re}}_{\rm {cr}} - {\rm Re_{ps}})^{-1/2}$ with $\dot{\gamma}$ being the shear rate. 
Huang \emph{et al}.~\cite{huang2012rotation} used the multiple-relation-time (MRT) model of LBM to numerically investigate spheroidal particles in simple shear flows in a wide range of ${\rm Re_{ps}}$ (from $0$ to $700$) and found several periodic and steady rotation modes. Moreover, recently, the dynamics of flexible fibers in turbulent flows have also been studied~\cite{olivieri2021universal,kunhappan2017numerical}. 
However, we find that the effects of inertia on the dynamics of flexible fibers particularly at finite Reynolds numbers have not been explored at all. 


In this work, we focus on the effects of fiber stiffness and confinement strength on the dynamics of single non-Brownian flexible fibers in 2D Couette flows (that are simple shear flows with uniform prescribed shear rates) at finite Reynolds numbers. For this purpose, we propose a new fiber-level simulation method by extending the FPD method for colloidal suspensions to study the dynamics of flexible fibers in viscous flows, where we have used a bead-spring fiber model that is similar to that proposed by Yamamoto and Matsuoka~\cite{yamamoto1993method}. Numerically, we implement the FPD method using the multiple-relaxation-time (MRT) scheme of LBM~\cite{mccracken2005multiple,huang2012rotation}. 
The FPD method, proposed by Tanaka and Araki~\cite{tanaka2000simulation}, was developed originally to deal with hydrodynamic interactions in suspensions of rigid colloids. This method approximates a solid particle as a highly viscous fluid and treats the fluid-solid interface as a diffuse interface, which not only avoids the explicit tracking of moving fluid-solid boundaries but they also reduce the computational cost significantly when simulating suspensions of a large number of particles.
\textcolor{black}{Moreover, the FPD method can be further extended to study the dynamics of Brownian fibers in viscous flows and fibers in complex fluids such as multiphase flows and fluids with an internal degree of freedom. Therefore, in comparison to other simulation methods, the FPD method is easier to implement and applicable to fiber dynamics in structured fluids with thermal fluctuations under confinement in semidilute or dense limits at finite Reynolds numbers.}
The LBM has provided an alternative and promising numerical scheme for simulating fluid flows and modeling physics in fluids~\cite{chen1998lattice}. In comparison to conventional numerical schemes based on discretizations of macroscopic continuum equations, the LBM is based on microscopic models and mesoscopic kinetic equations, which give LBM many of the advantages of molecular dynamics, including clear physical pictures, easy implementation, and fully parallel algorithms. 



The rest of the paper is organized as follows. In Sec.~\ref{Sec:Method}, the FPD method is introduced first in its original form for rigid particles in viscous \textcolor{black}{fluids}. We then extend it to study the dynamics of fibers in viscous \textcolor{black}{fluids} by using the bead-spring model of flexible fibers that have elastic resistance to both longitudinal compression/extensions and lateral bending. After that, we explain the numerical implementation of the FPD method using the MRT model of LBM. In Sec.~\ref{Sec:Benchmark}, the FPD method and the MRT-LBM numerical scheme are validated by two sets of Benchmark simulations that involve \textcolor{black}{fluid}-solid couplings in different geometries. In Secs.~\ref{Sec:Flexible}, our numerical method is then used to study the dynamics of flexible fibers in Couette flows with a focus placed on the effects of fiber stiffness, confinement strength, and Reynolds number. The paper is concluded in Sec.~\ref{Sec:Conclusions} with a few remarks. 



\section{Numerical method} \label{Sec:Method} 

In this section, we introduce the numerical methods adopted in this work to study the dynamics of flexible fibers in viscous flows that involve moving fluid-fiber interfaces and complex long-range hydrodynamic interactions. Firstly, we briefly explain the fluid particle dynamics (FPD) method that is originally developed for rigid particles in viscous flows. Secondly, we extend the FPD method to study flexible fibers in viscous \textcolor{black}{fluids} by using the bead-spring model where the fiber is regarded as a chain of connected rigid particles. Finally, we present the numerical implementation of the extended FPD method by using a multiple–relaxation–time (MRT) scheme of the lattice Boltzmann method (LBM).  

 
\subsection{Fluid Particle Dynamics (FPD) Method} \label{Sec:Method-FPD}

We first explain the FPD method briefly~\cite{tanaka2000simulation}. Consider a particle suspension with $N$ rigid particles immersed in a simple Newtonian \textcolor{black}{fluid} with viscosity $\eta_ {\rm f}$. In the FPD method, the suspension is treated as an incompressible viscous fluid with a spatially varying viscosity $\eta(\boldsymbol{r})$, which changes smoothly from the viscosity $\eta_ {\rm f}$ in the \textcolor{black}{fluids} to the large viscosity $\eta_ {\rm s}$ ($\gg \eta_ {\rm f}$) inside the rigid particle. The smooth viscosity profile $\eta(\boldsymbol{r})$ is usually chosen by expressing $\eta$ as a function of an interfacial profile function $\phi_i(\boldsymbol{r})$ that changes smoothly from $\phi_i=0$ in the \textcolor{black}{fluids} to $\phi_i=1$ inside the rigid particle. That is, the \textcolor{black}{fluid}-particle interface is treated as a diffusive interface with some finite thickness; this avoids the explicit tracking of the moving fluid-solid interfaces. To reproduce the no-slip boundary conditions at the solid surfaces of rigid particles, a linear monotonic function $\eta(\phi_i)$ is usually used~\cite{tanaka2000simulation,tanaka2006viscoelastic}:
\begin{equation}\label{Eq:Method-etaphi}
\eta(\boldsymbol{r})=\eta_{\rm f}+\sum_{i=1}^{N}(\eta_ {\rm s}-\eta_ {\rm f})\phi_i(\boldsymbol{r}). 
\end{equation}
However, if fluid slip is important at the particle surfaces, we have to use a non-monotonic function $\eta(\phi_i)$, in which an intermediate thin layer of lubricant with a very small viscosity has to be included \cite{zhang2015anisotropic}. 
For circular or spherical rigid particles~\cite{zhang2015anisotropic}, the most widely used form of $\phi_i(\boldsymbol{r})$ is
\begin{equation}\label{Eq:Method-phi}
    \phi_i(\boldsymbol{r}) = \frac{1}{2}\left\{ \tanh\left[\frac{1}{{\xi_i}}\left({a_i-|\boldsymbol{r}-\boldsymbol{R}_i|}\right)\right]+1 \right\},
\end{equation}
in which $\boldsymbol{R}_i$ is the center-of-mass position of the rigid particle $i$; $a_i$ and $\xi_i$ are the particle radius and the \textcolor{black}{fluid}-solid interface thickness, respectively. For later use, we denote particle diameters by $d_i=2a_i$. Other forms of the interfacial profile function $\phi_i$ can also be used, for example, the simple piece-wise linear function. 
We would like to point out that the interfacial profile function $\phi_i(\boldsymbol{r})$ defined in Eq.~(\ref{Eq:Method-phi}) is similar to but different from the level-set function in the level-set method~\cite{xu2014level} and the phase-field function in the phase-field method~\cite{liu2017fluid}. Here $\phi_i(\boldsymbol{r})$ at each time $t$ is not solved from dynamic equations but is prescribed mandatorily for a given center-of-mass position $\boldsymbol{R}_i(t)$ of the rigid particle that can change with time.

The dynamics of the incompressible viscous ``suspension" fluid characterized by the smooth viscosity profile $\eta({\boldsymbol{r}})$ is described by the Navier-Stokes equation:
\begin{equation} \label{Eq:Method-NS}
\rho\left(\frac{\partial \boldsymbol{v}}{\partial t}+\boldsymbol{v} \cdot \nabla \boldsymbol{v}\right)=-\nabla p+\nabla \cdot\left[\eta(\boldsymbol{r})\left(\nabla \boldsymbol{v}+\nabla \boldsymbol{v}^{T}\right)\right]+ \boldsymbol{f},
\end{equation}
and the incompressibility condition $\nabla \cdot \boldsymbol{v}=0$. 
Here $\rho$ is the mass density of the fluid, $\boldsymbol{v}(\boldsymbol{r})$ is the velocity field, $p$ is the pressure, and $\boldsymbol{f}$ is the force density due to external forces and/or the interactions between rigid particles, which is obtained from the forces $\boldsymbol{F}_i$ applied on the center-of-mass of each rigid particle by 
\begin{equation}\label{Eq:Method-force}
\boldsymbol{f}(\boldsymbol{r}) = \sum_{i=1}^N  
\frac{\phi_{i} \boldsymbol{F}_{i}}{\int d\boldsymbol{r} \phi_i}.
\end{equation}
The velocity of the rigid particle is given by 
\begin{equation}\label{Eq:Method-Vi}
\frac{d\boldsymbol{R}_i}{dt}=\boldsymbol{V}_i \equiv \frac{\int d\boldsymbol{r} \phi_i \boldsymbol{v}}{\int d\boldsymbol{r} \phi_i}.   
\end{equation} 

\noindent We would like to give several remarks on the above FPD method as follows. 

(1) In the FPD method, the solid region modeled by a highly viscous \textcolor{black}{fluid} can be regarded as rigid requiring the following two conditions: the solid-to-\textcolor{black}{fluid} viscosity ratio $r_\eta=\eta_{\rm s}/ \eta_{\rm f} \to \infty$ and the relative interfacial thickness $r_\xi=\xi/\ell \to 0$ with $\ell$ being a characteristic length of the system. \textcolor{black}{It has been shown by theory and numerical simulations that with increasing $r_\eta$ and decreasing $r_\xi$, the FPD approximation method is expected to become asymptotically exact~\cite{tanaka2006viscoelastic}. On the one hand, an infinite $r_{\eta}$ ensures that the shear-rate tensor is uniformly small within the solid region. Recently, similar ideas of imposing the constraint of zero shear-rate tensors in the solid region have been used to extend the diffuse interface models of two-phase flows to the fluid-solid coupling dynamics~\cite{hong2021hybrid}. On the other hand, an infinitesimal $r_\xi$ makes the interfacial region relatively small to mimic the sharp fluid-solid interface where fluids do not slip and the interface is impermeable. However, in practice, we have to make a trade-off between accuracy and computational cost. The specific values of $r_\eta$ and $r_\xi$ are usually taken empirically and depend on the requirement of computational accuracy. In most FPD simulations of colloidal suspensions~\cite{tanaka2000simulation,tanaka2006viscoelastic,tanaka2018physical}, one usually chooses $r_\eta \approx 50$ and $r_\xi\approx 0.1$. Here we will also carry out some benchmark simulations to find proper values of $r_{\eta}$ (and $r_\xi$) that should be large (small) enough to ensure accuracy but not be too large (small) to save computational cost. }

(2) Here we have only considered neutrally buoyant rigid particles or fibers. That is, we have neglected the difference between the \textcolor{black}{fluid} density and that of particles. More generally, one can add non-zero gravitational force in $\boldsymbol{f}$. Moreover, external torques can also be applied to each particle in addition to forces $\boldsymbol{F}_i=\rho \phi_i \boldsymbol{g}$. 

(3) To better see how the FPD method can describe the particle-\textcolor{black}{fluid} interactions, we can multiply both sides of Eq.~(\ref{Eq:Method-NS}) by $\phi_i(\boldsymbol{r})$, integrate over space, and obtain an approximate equation for the translational motion of rigid particles as~\cite{tanaka2018physical}
\begin{equation}\label{Eq:Method-ParticleEqn}
M_{i} {d \boldsymbol{V}_{i}}/{d t}=\boldsymbol{F}_{i}+\boldsymbol{F}_{{\rm v},i},
\end{equation} 
with $M_i$ and $\boldsymbol{V}_i$ being the mass and the velocity of particle $i$, respectively. Here $\boldsymbol{F}_{{\rm v},i}$ is the force exerted by the surrounding \textcolor{black}{fluids} on the rigid particle~\cite{goto2015purely}:
\begin{equation}\label{Eq:Method-Fv}
\boldsymbol{F}_{{\rm v},i}\approx \int d A_i \hat{\boldsymbol{n}}_{{\rm p},i} \cdot \left [ -p \boldsymbol{I}+ \eta_{\rm f} \left(\nabla \boldsymbol{v}+\nabla \boldsymbol{v}^{T}\right)\right].    
\end{equation}
in which $\boldsymbol{I}$ is the unit tensor, the integral is over the particle surface, $dA_i$ and $\hat{\boldsymbol{n}}_{{\rm p},i}$ are the surface element and the unit outward normal vector of the particle, respectively. 

Particularly, we are interested in this work in particle motion under the overdamped limits, where the characteristic time scaling as $d/V_0$ is much larger than the viscous relaxation time $M/\eta_{\rm f} d\approx \rho d^2/\eta_{\rm f}$ with $M$, $d=2a$, $a$, $V_0$ being the mass, diameter, radius, and characteristic velocity of the particle. That is, the particle Reynolds number, defined by
\begin{equation}\label{Eq:Method-Rep}
{\rm Re}_{\rm p}\equiv \rho d V_0/\eta_{\rm f},  
\end{equation}
is very small: ${\rm Re}_{\rm p} \ll 1$. In this limit, we have $\boldsymbol{F}_{i}+\boldsymbol{F}_{{\rm v},i}=0$ and the center-of-mass velocities $\boldsymbol{V}_i$ of rigid particles are constants, independent of time. \textcolor{black}{Furthermore, for particles or fibers moving in viscous flows at small Reynolds numbers and weak confinement, the total force $\boldsymbol{F}_{{\rm v},i}$ applied by fluids on the particles can be simply divided into the drag force due to the relative particle-fluids motion and the two-body lubrication force between particles when they get very close to each other, as used for example in Ref.~[50]. 
Such treatment simplifies the dynamic simulations of single flexible fiber and their suspensions significantly. However, for more general cases of finite Reynolds numbers and strong confinement strength (or in dense fiber suspensions) as focused on in this work, the distinction between drag and lubrication forces is not practical because the forces applied by fluids on the particles or fibers show a strong dependence on the Reynolds number and involve many-body hydrodynamic interactions. This further necessitates the use of the numerical method presented in this work to study fiber dynamics in viscous flows under strong confinement (or in dense fiber suspensions) at finite Reynolds numbers. Of course, a comparative study using these different numerical methods will be important to find their validity and limitations in simulating the dynamics of both single fibers and fiber suspensions. } 

\begin{figure*} 
    \centering
	\includegraphics[width=1.8\columnwidth]{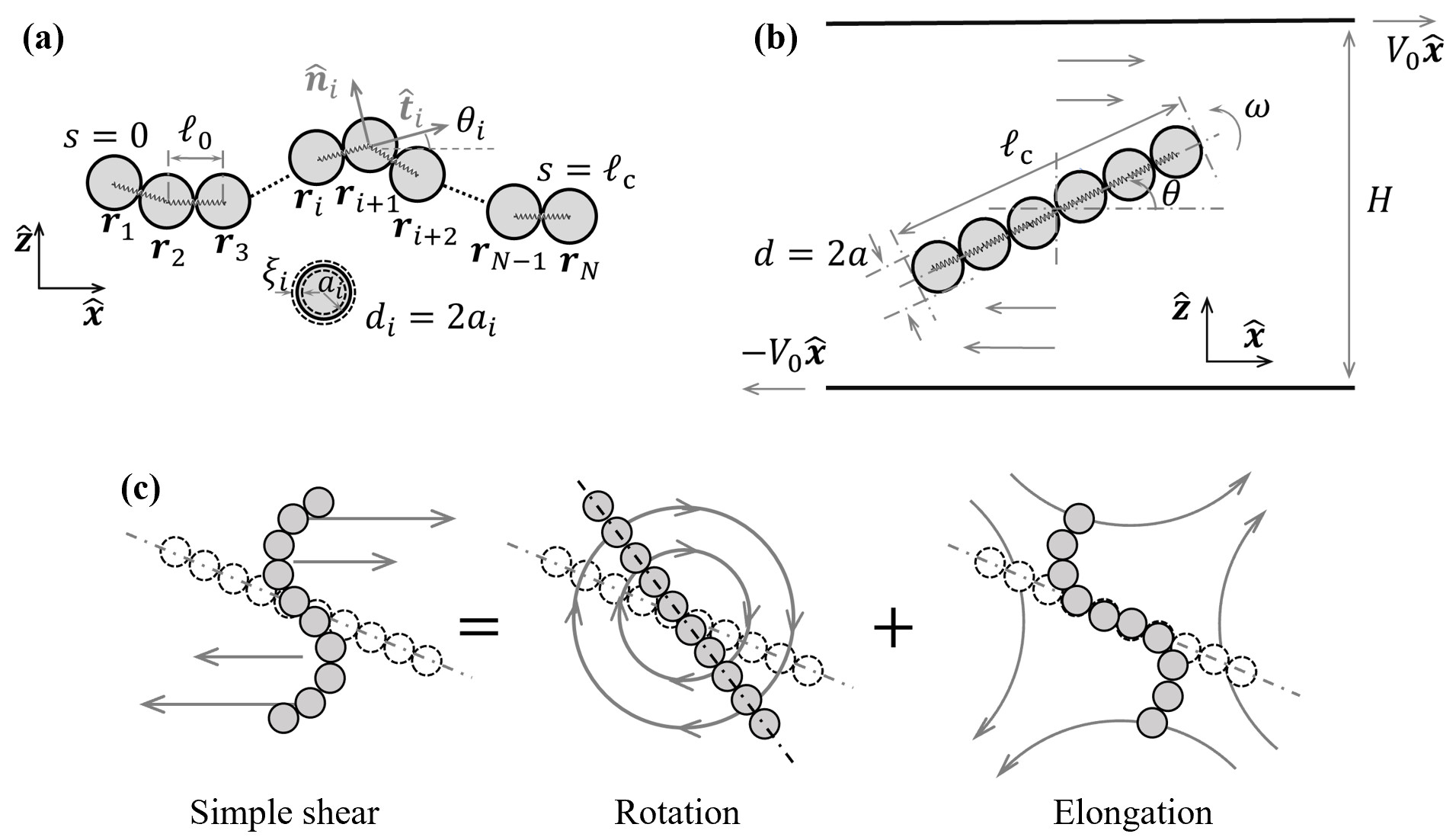}
	\caption{Schematic illustrations of (a) the bead-spring model for the non-Brownian flexible fibers and (b) the computational domain for a flexible fiber in a two-dimensional simple-shear flow that is confined between two solid walls. In (a), $d_i=2a_i$, $a_i$, $\xi_i$, and $\boldsymbol{R}_i$ are the diameter, the radius, the interfacial thickness, and the position vector of each rigid particle of the fiber, respectively. $\hat{\boldsymbol{n}}_i$ and  $\hat{\boldsymbol{t}}_i$ are the normal and tangential unit vectors of the fiber, respectively. $\theta_{i}$ denotes the angle of $\hat{\boldsymbol{t}}_i$ relative to the horizontal $+\hat{\boldsymbol{x}}$-direction. In (b), the velocities of the top and bottom walls are $\pm V_0$, respectively, and the distance between the two walls is $H$ and \textcolor{black}{the horizontal $x$-dimension is $L_{\rm x}$.}. Hence, the shear rate is $\dot{\gamma}=2V_0/H$. $d=2a$ and $a$ are the diameter and radius of each rigid particle of the fiber, respectively. $\omega$ is the angular velocity of the fiber, and $\omega$ is a function of fiber orientation $\theta$ relative to the horizontal $+\hat{\boldsymbol{x}}$-direction. Here $\ell_{\rm c}$ denotes the contour length of the fiber and the aspect ratio is $r_{\rm c}=d/\ell_{\rm c}$. (c) Decomposition of simple shear flow into the rigid-body rotation and the elongation (pure shear) flow. Consequently, the flexible fiber centered in the simple shear flow  tumble (or rotate) and deform (or buckles).}
    \label{Fig:Method-Schematic}
\end{figure*}


\subsection{Bead-spring model for flexible fibers} \label{Sec:Method-Fiber}

Next, we use the bead-spring model of flexible fibers~\cite{yamamoto1993method} to extend the FPD method for rigid particles to study the dynamics of flexible fibers in viscous \textcolor{black}{fluids}. A similar idea has been used to study the rheology of dilute suspensions of non-Brownian elastic dumbbells that are composed of two rigid particles connected by a Hookean spring~\cite{peyla2007rheology}. As shown schematically in Fig.~\ref{Fig:Method-Schematic}(a), the fiber consists of $N$ identical rigid circular beads (of radius, $a$, and thickness, $\xi$) that are connected by ($N-1$) identical springs (of equilibrium length $\ell_0=d=2a$). Flexible fibers show elastic resistance to both compression and bending, and hence we take the total energy (per length) in 2D to be the following form
\begin{equation}\label{Eq:Method-U}
U\left(\boldsymbol{R}_i\right)=\sum_{i=1}^{N-1}\frac{k_{\rm s}}{2{\ell_0}^2}\left(R_{i,i+1}-\ell_0\right)^{2} -\sum_{i=1}^{N-1}k_{\rm b}\left(\boldsymbol{\hat{t}}_{i+1}\cdot \boldsymbol{\hat{t}}_{i}\right)^{2}.
\end{equation}
Here $R_{i,i+1}=|\boldsymbol{R}_{i,i+1}|$ is the distance between the center-of-mass of the two neighboring beads with $\boldsymbol{R}_{i,i+1}=\boldsymbol{R}_{i+1}-\boldsymbol{R}_{i}$ and $\boldsymbol{R}_i$ being the center-of-mass position of each bead. $\hat{\boldsymbol{t}}_{i}=\boldsymbol{R}_{i,i+1}/R_{i,i+1}$ is tangential unit vector of the fiber. $k_{\rm {s}}$ is the spring constant, and for inextensible fibers, a very large $k_{\rm {s}}$ should be used such that the spring length is close to its equilibrium length $\ell_0$, and hence the contour length of the fiber is also almost constant to be $\ell_{\rm c}=N\ell_0$. $k_{\rm b}$ is the bending constant that characterizes the stiffness of the fiber. The total force $\boldsymbol{F}_i$ (per length) acting on each rigid bead of the non-Brownian flexible fiber is then given by $\boldsymbol{F}_i=-{\partial U}/{\partial\boldsymbol{R}_i}$, which include two parts: the compression force $\boldsymbol{F}_{i}^{\rm {s}}$ and the bending force $\boldsymbol{F}_{i}^{\rm {b}}$ due to the changes in the distance between neighboring beads and in the local orientation (bending), respectively,  that is,
\begin{equation}\label{Eq:Method-BeadSpringFi}
\boldsymbol{F}_i=\boldsymbol{F}_{i}^{\rm {s }}+\boldsymbol{F}_{i}^{\rm {b}}.
\end{equation} 
Firstly, from the first term of the total energy in Eq.~(\ref{Eq:Method-U}), we obtain the compression force $\boldsymbol{F}_{i}^{\rm {s}}$ for $2<i<N-1$:
\begin{subequations}\label{Eq:Method-BeadSpringFs}
\begin{equation}\label{Eq:Method-BeadSpringFsi}
\boldsymbol{F}_{i}^{\rm {s }}=-\frac{k_{s}}{\ell_0^2}\left[\left(R_{i-1, i}-\ell_0\right) \hat{\boldsymbol{t}}_{i-1}-\left(R_{i, i+1}-\ell_0\right) \hat{\boldsymbol{t}}_{i}\right],
\end{equation}
and for the two beads at the two ends:
\begin{equation}\label{Eq:Method-BeadSpringFs1N}
\boldsymbol{F}_{1}^{\rm {s}}=\frac{k_{s}}{\ell_0^2}\left(R_{1,2}-\ell_0\right) \hat{\boldsymbol{t}}_{1},\quad \boldsymbol{F}_{N}^{\rm {s}}=-\frac{k_{s}}{\ell_0^2}\left(R_{N-1, N}-\ell_0\right) \hat{\boldsymbol{t}}_{N-1}.
\end{equation}
\end{subequations}
Secondly, from the second term of the total energy in Eq.~(\ref{Eq:Method-U}), we obtain the bending force $\boldsymbol{F}_{i}^{\rm {b}}$ for $3<i<N-2$:
\begin{subequations}\label{Eq:Method-BeadSpringFb}
\begin{align}\label{Eq:Method-BeadSpringFbi}
\boldsymbol{F}_{i}^{\rm {b}}= k_{b}\left[\left(\hat{\boldsymbol{t}}_{i}+\hat{\boldsymbol{t}}_{i-2}\right) \cdot \frac{\left(\boldsymbol{I}-\hat{\boldsymbol{t}}_{i-1} \hat{\boldsymbol{t}}_{i-1}\right)}{R_{i-1, i}} \right.\\
\nonumber 
\left. -\left(\hat{\boldsymbol{t}}_{i+1}+\hat{\boldsymbol{t}}_{i-1}\right) \cdot \frac{\left(\boldsymbol{I}-\hat{\boldsymbol{t}}_{i} \hat{\boldsymbol{t}}_{i}\right)}{R_{i, i+1}}\right],
\end{align}
with $\boldsymbol{I}$ being the unit tensor, and for edge beads (the last two beads at the two ends):
\begin{equation}\label{Eq:Method-BeadSpringFb1N}
\boldsymbol{F}_{1}^{\rm {b }}=-k_{b} \hat{\boldsymbol{t}}_{2} \cdot \frac{\left(\boldsymbol{I}-\hat{\boldsymbol{t}}_{1} \hat{\boldsymbol{t}}_{1}\right)}{R_{1,2}},
\end{equation}
\begin{equation}\label{Eq:Method-BeadSpringFb1N2}
\boldsymbol{F}_{2}^{\rm {b }}=k_{b}\left[\hat{\boldsymbol{t}}_{2} \cdot \frac{\left(\boldsymbol{I}-\hat{\boldsymbol{t}}_{1} \hat{\boldsymbol{t}}_{1}\right)}{R_{1,2}}-\left(\hat{\boldsymbol{t}}_{3}+\hat{\boldsymbol{t}}_{1}\right) \cdot \frac{\left(\boldsymbol{I}-\hat{\boldsymbol{t}}_{2} \hat{\boldsymbol{t}}_{2}\right)}{R_{2,3}}\right],
\end{equation}
\begin{align}\label{Eq:Method-BeadSpringFbN1N3}
 \boldsymbol{F}_{N-1}^{\mathrm{b}}=k_{b}\left[\left(\hat{\boldsymbol{t}}_{N-1}+\hat{\boldsymbol{t}}_{N-3}\right) \cdot \frac{\left(\boldsymbol{I}-\hat{\boldsymbol{t}}_{N-2} \hat{\boldsymbol{t}}_{\boldsymbol{N}-2}\right)}{R_{N-2, N-1}}\right.
 \\  \nonumber
 \left.-\hat{\boldsymbol{t}}_{N-2} \cdot \frac{\left(\boldsymbol{I}-\hat{\boldsymbol{t}}_{N-1} \hat{\boldsymbol{t}}_{N-1}\right)}{R_{N-1, N}}\right],   
\end{align}
\begin{equation}\label{Eq:Method-BeadSpringFbN1N4} 
\boldsymbol{F}_{N}^{\rm {b}}=k_{b} \hat{\boldsymbol{t}}_{N-2} \cdot \frac{\left(\boldsymbol{I}-\hat{\boldsymbol{t}}_{N-1} \hat{\boldsymbol{t}}_{N-1}\right)}{R_{N-1, N}}. 
\end{equation} 
\textcolor{black}{Similar energy and forces have also been used and derived in Ref. [24]
and in recent works in Refs. [3,23]. }

\end{subequations} 

\subsection{Numerical solution method: Lattice Boltzmann method (LBM)} \label{Sec:Method-LBM}

We present our numerical algorithm for solving the dynamic equations (\ref{Eq:Method-etaphi})--(\ref{Eq:Method-Vi}) with the forces $\boldsymbol{F}_i$ in Eqs.~(\ref{Eq:Method-BeadSpringFi})--(\ref{Eq:Method-BeadSpringFb}) applied on each fiber bead. Firstly, to obtain a set of dimensionless equations suitable for numerical computations, we scale the length by the particle diameter $d=2a$, velocity by a characteristic velocity $V_0$, time by $d/V_0$, pressure or stress by $\eta_{\rm f} V_0/d$, and force by $\eta_{\rm f} V_0$. Five dimensionless parameters appear as follows. (1) The particle Reynolds number: ${\rm Re}_{\rm p}\equiv \rho d V_0/\eta_{\rm f}$ as defined in Eq.~(\ref{Eq:Method-Rep}). (2) The compression stiffness parameter of the fiber: $\mathcal{K}_{\rm s}={k_ {\rm{s}}}/{\eta_{\rm f} V_0 \ell_0}$ (with $\ell_0=d$). (3) The bending stiffness parameter of the fiber:  $\mathcal{K}_{\rm b}={k_{\rm b}}/{\eta_{\rm f} V_0 \ell_0}$. (4) The particle-to-\textcolor{black}{fluid} viscosity ratio: $r_{{\eta}}\equiv \eta_ {\rm s}/\eta_ {\rm f}\gg 1$. (5) The relative particle-\textcolor{black}{fluid} interface thickness: $r_{\xi}\equiv {\xi}/d$.  
\textcolor{black}{In our simulations, we only consider the dynamics of inextensible flexible fibers where the compression stiffness parameter ${\mathcal K}_{\rm s}$ is set to be a very large value, ${\mathcal K}_{\rm s}=3.33\times 10^{6}$. In this case, the spring length (or the neighbor particle-particle distance) is kept to be around its equilibrium length $\ell_0\approx d$ and hence no gaps between neighbor rigid particles form in the fiber and the fiber is impermeable for fluids. }

In this work, the Navier-Stokes (NS) equation (\ref{Eq:Method-NS}) and the dynamic equation (\ref{Eq:Method-Vi}) for the center-of-mass position of rigid particles are numerically solved by the lattice Boltzmann method (LBM) and by the forward finite difference method, respectively. LBM is a promising simulation technique that has attracted interest from researchers in computational physics~\cite{chen1998lattice}. LBM has many advantages over other conventional computational fluid dynamics methods, especially in dealing with interfacial phenomena.  
The conventional LBM algorithm for NS equation with constant viscosity usually uses a single-relaxation-time Bhatnagar-Gross-Krook (BGK) collision operator~\cite{chen1998lattice}. However, our NS equation involves a large viscosity ratio; to improve the numerical accuracy and stability, we employ the multiple-relaxation-time (MRT) collision model for LBM~\cite{mccracken2005multiple,huang2012rotation}, the centered formulation for the forcing term~\cite{chen1998lattice}, and the isotropic discretization based on the D2Q9 velocity model~\cite{chen1998lattice}) to evaluate the spatial gradients of the phase-field variables. The collision operators by the MRT method have more degrees of freedom than the BGK operator, which can be used to improve accuracy and stability~\cite{chen1998lattice}. Details of the LBM method and the MRT collision model are well described in the literature such as in Refs. [41,44];
for conciseness, we will not repeat the formulations.  

In the numerical simulations, the spatial domain is rectangular, specified by $0 \leq x \leq L_x$, $0 \leq z \leq H$, and this domain is discretized into $N_x \times N_z$ uniform lattice squares of step length $\Delta x$, giving $L_x = N_x\Delta x$ and $H = N_z\Delta x$. We place the following fluid variables at the center of the discrete lattice squares (\emph{i.e.}, on-lattice variables): the distribution functions in LBM, the velocity field $\boldsymbol{v}(\boldsymbol{r})$, the pressure field $p(\boldsymbol{r})$, the interfacial profile function $\phi_i(\boldsymbol{r})$, and the viscosity profile function $\eta(\boldsymbol{r})$. In contrast, the center-of-mass position $\boldsymbol{R}_i$ of the rigid particle is off-lattice. 

Before ending this section, we outline the integration steps for implementing the FPD method using MRT-LBM as follows. 
We want to solve the spatiotemporal evolution of the velocity field $\boldsymbol{v}(\boldsymbol{r},t)$ and the temporal evolution of the center-of-mass position $\boldsymbol{R}_i(t)$ of each fiber particle from their initial conditions. The superscript $n$ denotes consecutive time instants and $\Delta t$ is the time interval.
\begin{itemize}
\item [(1)] At time instant $n$, compute the interfacial profile $\phi_i^n(\boldsymbol{r})$, the force density field $\boldsymbol{f}^n (\boldsymbol{r})$, and the viscosity profile $\eta^n(\boldsymbol{r})$. 

We first use the center-of-mass position $\boldsymbol{R}_i^n$ of each fiber particle to generate $\phi_i^n(\boldsymbol{r})$ according to Eq.~(\ref{Eq:Method-phi}), compute the forces $\boldsymbol{F}^n_i$ on each fiber particle from Eqs.~(\ref{Eq:Method-BeadSpringFi})--(\ref{Eq:Method-BeadSpringFb}), and $\boldsymbol{f}^n(\boldsymbol{r})$ from Eq.~(\ref{Eq:Method-force}). Then, we use $\phi_i^n(\boldsymbol{r})$ and Eq.~(\ref{Eq:Method-etaphi}) to compute $\eta^n(\boldsymbol{r})$.  

\item [(2)] Use the MRT-LBM algorithm to solve the velocity field $\boldsymbol{v}^{n+1}(\boldsymbol{r})$ from NS equation (\ref{Eq:Method-NS}) using the on-lattice fields, $\boldsymbol{v}^{n}(\boldsymbol{r})$, $\phi_i^n(\boldsymbol{r})$, $\boldsymbol{f}^n (\boldsymbol{r})$, and $\eta^n(\boldsymbol{r})$ at time instant $n$.

\item [(3)] Compute the velocities $\boldsymbol{V}_i^{n+1}$ of each particle using Eq.~(\ref{Eq:Method-Vi}) and the updated the velocity field $\boldsymbol{v}^{n+1}(\boldsymbol{r})$. Then we update the off-lattice center-of-mass positions of rigid particles by $\boldsymbol{R}_i^{n+1}=\boldsymbol{R}_i^n+\Delta t \boldsymbol{V}_i^{n+1}$. 
\end{itemize}


\section{Benchmark simulations of fluid-solid coupling} \label{Sec:Benchmark}

\begin{figure*} [htb] 
    \centering
	\includegraphics[width=2\columnwidth]{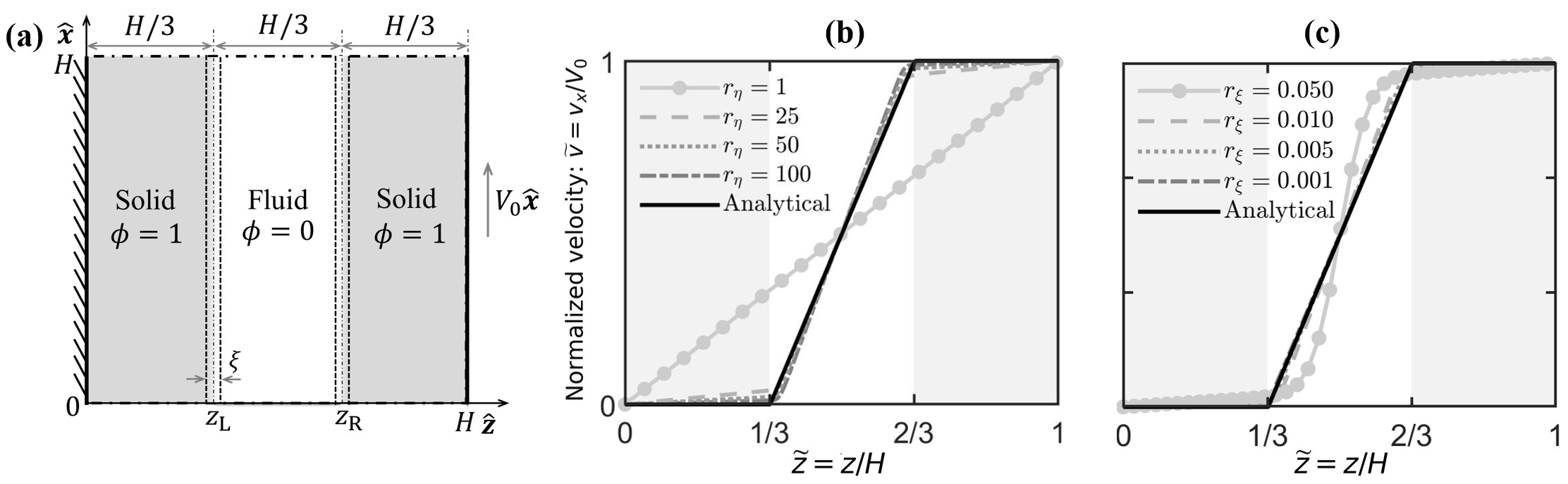}
	\caption{Benchmark simulation 1: Couette flows (confined simple shear flows). (a) Schematic illustration of the sandwiched (solid-\textcolor{black}{fluid}-solid) computational domain that generates Couette flows. The velocities at the left (at $z=0$) and right (at $z=H$) boundaries to be $0$ and $V_0\hat{\boldsymbol{x}}$, respectively. The solid region is represented by a highly viscous \textcolor{black}{fluid} with viscosity $\eta_{\rm s}$ that is much larger than \textcolor{black}{fluid} viscosity $\eta_{\rm f}$. (b,c) Profile of lateral velocity $v_x(z)$ for different viscosity ratios $r_\eta=\eta_{\rm s}/\eta_{\rm f}$ and relative thicknesses $r_\xi=\xi/H$. In (b), we fix $r_\xi=0.01$, and in (c) we fix $r_\eta=100$. The analytical solution in Eq.~(\ref{eq:Benchmark-Couette-Analytical}) is shown by the black solid lines.} 
    \label{Fig:Benchmark1}
\end{figure*}

\begin{figure*} [!htb] 
    \centering
	\includegraphics[width=1.75\columnwidth]{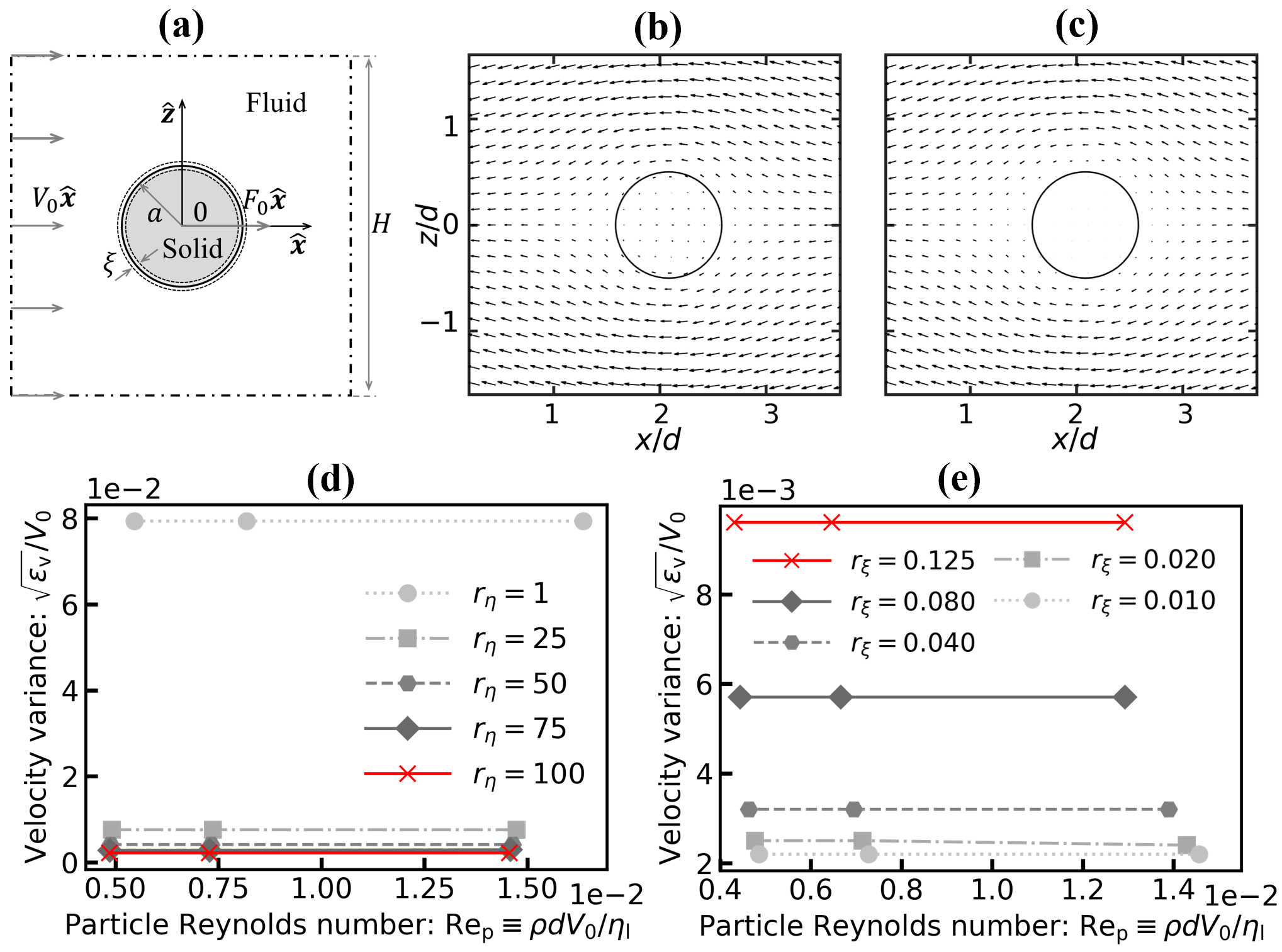}
	\caption{Benchmark simulation 2: A circular rigid particle moving in viscous \textcolor{black}{fluids}. (a) Schematic illustration of a circular particle of radius $a$ immersed in viscous \textcolor{black}{fluids}. The following two equivalent steady-state dynamics are studied. (1) The steady-state motion of the particle with velocity $\boldsymbol{V}_0$ driven by a constant external force ${\boldsymbol F}_0 =F_0 \hat{\boldsymbol{x}}$, where the surrounding \textcolor{black}{fluid} is stationary far from the particle. (2) The steady flow with a uniform inlet velocity of $\boldsymbol{V}_0$ passing the particle fixed in the center of the domain. \textcolor{black}{(b,c) Snapshots of the velocity field (arrows) near and inside the particle (circle) for $r_{\eta}=1$ and $r_{\eta}=100$, respectively. Here the whole computational domain is square and the domain size $H=10d$ with $d$ being the particle diameter, that is, the snapshots are only parts of the whole computational domain. } (d,e) Dependence of the strength of flows inside particles for various viscosity ratios $r_\eta$, the relative thickness of interface $r_\xi$, and the particle Reynolds number $\rm Re_p$. Here we set $r_{\xi}=0.01$ in (b,c,d), and $r_{\eta}=100$ in (e). }
    \label{Fig:Benchmark2}
\end{figure*}

We first carry out two sets of benchmark simulations involving \textcolor{black}{fluid}-solid coupling to check the validity of the FPD method that is implemented using the MRT-LBM algorithm. (1) Couette flows or simple shear flows between two solid walls moving at different velocities, where we calculate the velocity profile. (2) A circular particle moving in viscous \textcolor{black}{fluids}, where we calculate the drag forces for different Reynolds numbers. Our numerical results are then compared with analytical results or numerical results obtained by other methods in the literature. 

\subsection{Couette flows: simple shear flows between two moving solid walls}\label{Sec:Benchmark-Couette}

Couette flow is a typical shear flow of viscous fluids between two parallel solid walls moving at different velocities. Here we generate such a flow in a sandwich (solid-\textcolor{black}{fluid}-solid) system as shown in Fig.~\ref{Fig:Benchmark1}(a) by setting the velocities at the left (at $z=0$) and right (at $z=H$) boundaries to be $0$ and $V_0\hat{\boldsymbol{x}}$, respectively. In this system, two \textcolor{black}{fluid}-solid interfaces appear at $z_{\rm f}=H/3$ and $z_{\rm R}=2H/3$, respectively, where simple \textcolor{black}{fluid}-solid coupling appears, the no-slip and impenetrable boundary conditions apply. In the lateral $x$-direction, the periodic boundary condition is employed. In this case, the analytical solution of the lateral velocity $v_x$ can be obtained easily~\cite{Landau2013fluid} as  
\begin{equation}\label{eq:Benchmark-Couette-Analytical}
v_x =
  \begin{cases}
  0, & 0 \leq z < z_{\rm f}, \\
  V_0 (z/z_{\rm f}-1), & z_{\rm f} \leq  z \leq z_{\rm R}, \\
  V_0, & z_{\rm R} <z<H.
  \end{cases}
\end{equation}    
Following the FPD method mentioned in Sec.~\ref{Sec:Method-FPD}, the sandwich system can be represented by the following continuous interfacial profile function:
\begin{equation}\label{Eq:Benchmark-phi} 
    \phi(z)=
    \begin{cases}
        \frac{1}{2}\left\{\tanh\left[(z_L-z)/\xi\right]+1\right\}, \quad  z\leq H/2, \\
        \frac{1}{2}\left\{\tanh\left[(z-z_R)/\xi\right]+1\right\},\quad z>H/2.
    \end{cases}
\end{equation}
Accordingly, the viscosity profile is given from Eq.~(\ref{Eq:Method-etaphi}) by $\eta(z)=\eta_{\rm f}+(\eta_ {\rm s}-\eta_ {\rm f})\phi(z)$. Then, in the \textcolor{black}{fluid} region, $\phi=0$ and the viscosity is $\eta_{\rm f}$. In the solid region, $\phi=1$ and the viscosity is $\eta_{\rm s}$ that should be set to be much larger than \textcolor{black}{fluid} viscosity $\eta_{\rm f}$.

In our numerical simulations, the spatial domain is square with $0 \leq x \leq H$ and $0 \leq z \leq H$; the normalized spatial step $\Delta x/H=\Delta z/H$ and time step $\Delta t/(H/V_0)$ are both given by $0.005$, respectively. The Reynolds number $\rm{Re}={\rho V_0 H}/{\eta_{\rm f}}$ is set as $0.1$. Moreover, to make the trade-off between accuracy and computational cost, we explore the effects of $r_\eta=\eta_{\rm s}/\eta_{\rm f}$ and $r_\xi=\xi/H$ on the accuracy of numerical results of velocity field $v_x$ as shown in Fig.~\ref{Fig:Benchmark1}(b,c). As $r_\eta$ increases and $r_\xi$ decreases, the numerical results fit the analytical solution in Eq.~(\ref{eq:Benchmark-Couette-Analytical}) better. Particularly, when $r_\eta\geq 50$ and $r_\xi\leq 0.01$, the numerical results have already fit the analytical solution very well. 

\subsection{A circular particle moving in viscous \textcolor{black}{fluids}: viscous drag forces}\label{Sec:Benchmark-Drag}

In the second set of benchmark simulations, we consider a circular rigid particle moving in two-dimensional viscous \textcolor{black}{fluids}, where the \textcolor{black}{fluid}-solid coupling is more directly related to that of fibers in viscous shear flows. 
As shown in Fig.~\ref {Fig:Benchmark2}(a), a circular rigid particle of radius $a$ is immersed in an unbounded viscous flow with the viscosity of $\eta_{\rm{l}}$. In the numerical simulations, we still take the square spatial domain with $0 \leq x \leq H$ and $0 \leq z \leq H$. The normalized spatial step $\Delta x/d=\Delta z/d$ and time step $\Delta t/(d/V_0)$ are both given by $0.05$, respectively. The normalized system size is chosen to be $H/d=10$ and periodic boundary conditions are employed at the boundaries of the computational domain. \textcolor{black}{The size $H=10d$ of the computational domain is large enough so that the artificial effects introduced by periodic boundary conditions are negligible. For example, for the simulations shown in Fig.~\ref{Fig:Benchmark2}(b,c), the velocity far away from the particle is too small to be visible (\emph{i.e.}, stationary as expected for the particle motion in infinite systems). }

We first investigate the steady-flow field around the circular particle, which is driven by a constant external force ${\boldsymbol F}_0 =F_0 \hat{\boldsymbol{x}}$ applied on the particle as shown in Fig.~\ref{Fig:Benchmark2}(a). Moreover, we assume the surrounding \textcolor{black}{fluids} are stationary in the far field away from the particle. As the computational time is larger than the viscous relaxation time $\rho d^2/\eta_{\rm f}$, the velocity field and the particle motion reach a steady state with particle velocity $\boldsymbol{V}_0=V_0 \hat{\boldsymbol{x}}$. 
The validity of the FPD method that models rigid particles by highly viscous fluids can be evaluated by the strength of the residual flow fields inside the particle (measured in the moving reference frame fixed on the particle), which is defined by
\begin{equation}
\varepsilon_{\rm v} = \frac{\int d\boldsymbol{r} \phi (\boldsymbol{v}-\boldsymbol{V}_0)^2}{\int d\boldsymbol{r} \phi}.
\end{equation}
As discussed above, the FPD method should be accurate in the limits of infinite viscosity ratio $r_\eta \to \infty$ and infinitesimal interface thickness $r_\xi \to 0$. However, in practical simulations, we need to find their proper values with both low computational costs and enough accuracy. As shown in Fig.~\ref{Fig:Benchmark2}(d,e), as $r_\eta=\eta_{\rm s}/\eta_{\rm f}$ increases and $r_\xi=\xi/d$ decreases, the strength of internal flow field $\varepsilon_{\rm v}$ decreases and shows a very weak dependence on the particle Reynolds number $\rm{Re_p}$ defined in Eq.~(\ref{Eq:Method-Rep}). Particularly, when $r_\eta\geq 50$ and $r_\xi\leq 0.125$, the FPD method is good enough and the fluid particle behaves more like a rigid solid particle.

\begin{figure} [!htb]
\centering 
\includegraphics[width=0.85\columnwidth]{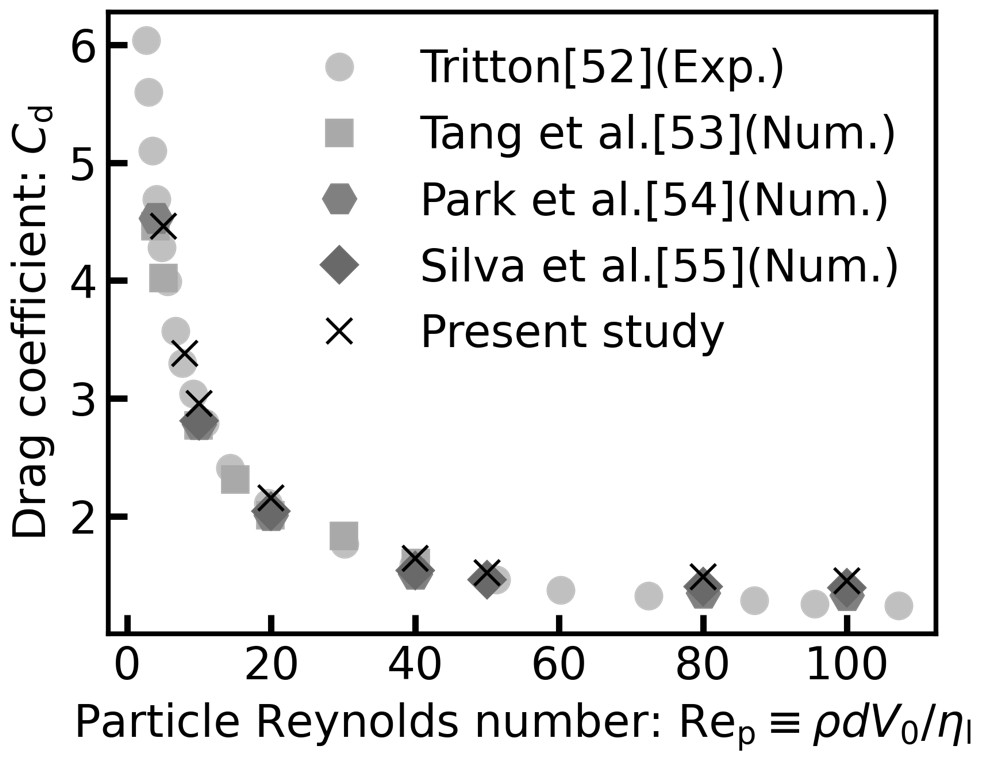}
\caption{Drag coefficient $C_{\rm d}$ plotted as a function of the particle Reynolds number ${ \rm {Re}_{\rm p}}={\rho v_0 d}/{\eta_l}$ defined in Eq.~(\ref{Eq:Method-Rep}). The viscosity ratio and thickness ratio are set as $r_\eta=\eta_{\rm s}/\eta_{\rm f}=50$ and $r_\xi=\xi/d= \textcolor{black}{0.0125}$, respectively.}
\label{Fig:Benchmark-CD}
\end{figure}

In addition, we also calculate the drag force (per length) $\boldsymbol{F}_{\rm d}$ applied by the surrounding viscous \textcolor{black}{fluids} on the particle. In the steady particle motion with velocity ${\boldsymbol V}_0$ driven by ${\boldsymbol F}_0$, we have $\boldsymbol{F}_{\rm d}=-{\boldsymbol F}_0$, and hence by varying ${\boldsymbol F}_0$ we can obtain $\boldsymbol{F}_{\rm d}$ as a function of ${\boldsymbol V}_0$ or $\rm {Re}_{\rm p}$, from which we can calculate the drag coefficient $C_{\rm d}$ by \textcolor{black}{
\begin{equation}\label{Eq:Benchmark-CD}
C_{\rm d}= \frac{|\boldsymbol{F}_{\rm d}|}{\rho {V_0}^{2} d/2}
= \frac{|\boldsymbol{F}_{\rm d}|/\eta_{\rm f}V_0}{{\rm {Re_p}/2}},
\end{equation} 
and since $C_{\rm d} \sim 1$, then the de-dimensionlized drage force $|\boldsymbol{F}_{\rm d}|/\eta_{\rm f}V_0$ scales as $|\boldsymbol{F}_{\rm d}|/\eta_{\rm f}V_0\sim {\rm Re_p}$.} However, in practical simulations, there is another more convenient setup to calculate $\boldsymbol{F}_{\rm d}$ as shown in Fig.~\ref{Fig:Benchmark2}(a): the steady flow with a uniform inlet velocity of $\boldsymbol{V}_0$ passing the particle fixed in the center of the domain. In this case, the drag force $\boldsymbol{F}_{\rm d}$ applied on the particle is given by
\begin{equation}\label{Eq:Benchmark-Fd}
\boldsymbol{F}_{\rm d}= - \int d A \hat{\boldsymbol{n}} \cdot \boldsymbol{\sigma}
=\frac{\int d\boldsymbol{r} \phi\hat{\boldsymbol{n}} \cdot \boldsymbol{\sigma}}{\int d\boldsymbol{r} \phi},
\end{equation}
where the first surface integral is over the particle surface, the second volume integral is over the domain with $\phi \neq 0$, $\boldsymbol{\sigma}=\eta (\nabla \boldsymbol{v}+\nabla \boldsymbol{v}^{\rm T})$ is the viscous stress tensor, and $\boldsymbol{n}$ is the outward unit normal vector of the particle. Using Eqs.~(\ref{Eq:Benchmark-CD}) and (\ref{Eq:Benchmark-Fd}), we calculate $C_{\rm d}$ as shown in Fig.~\ref{Fig:Benchmark-CD}, from which we see a very good agreement of $C_{\rm d}$ calculated using our numerical method with that measured in previous experiments by Tritton~\cite{tritton1959experiments}, and in numerical simulations by Tang~\emph{et al}.~\cite{tang2019investigation}, Park \emph{et al}.~\cite{park1998numerical}, and Silva \emph{et al}.~\cite{silva2003numerical}, for $\rm {Re}_{\rm p}$ from $10$ to $100$. These results confirm the validity of our numerical method to study the dynamics of solid particles in viscous \textcolor{black}{fluids}.

In this section, we have conducted two sets of benchmark simulations and have confirmed the validity of the FPD method implemented by MRE-LBM scheme in studying \textcolor{black}{fluid}-solid coupling problems. Practically, we find that the computational accuracy of our numerical simulations is high enough if $r_\eta=\eta_{\rm s}/\eta_{\rm f}\ge 50$ and $r_\xi=\xi/d \le 0.125$. Therefore, to ensure accuracy and save computational cost, we choose $r_\eta=100$ and $r_\xi=0.125$ in the simulations presented later in this work.

\section{Flexible fibers in Couette flows}\label{Sec:Flexible}

\subsection{Dimensionless parameters}\label{Sec:Flexible-Dimensionless}

\begin{figure*} [!htb]
\centering
\includegraphics[width=1.9\columnwidth]{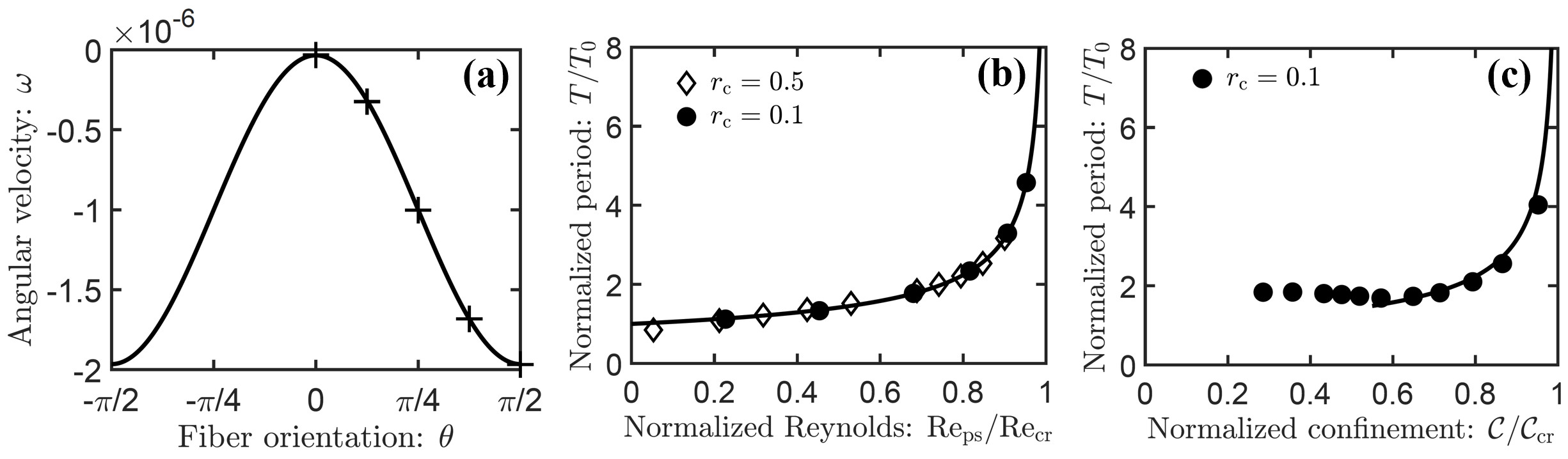}
\caption{\textcolor{black}{Tumbling dynamics of rigid fibers (with $\mathcal{K}=3.3 \times \textcolor{black}{10^0}$).
(a) Jeffery orbit of free rigid fibers with tumbling angular velocity $\omega$ following Eq.~(\ref{Eq:Flexible-omega}) with ${\rm{Re}_{ps}}\ll 1$ and $\mathcal{C}=0.2$. 
(b) Dependence of the tumbling period $T$ (normalized by the extrapolated period $T_0$ at ${\rm{Re}_{ps}}=0$) on ${\rm{Re}_{ps}}$. 
The period $T$ for fibers of different aspect ratios follows a unified power-law scaling (solid line): $\dot {\gamma} T \propto ({\rm {Re}}_{\rm {cr}} -{\rm{Re}_{\rm ps}})^{-1/2}$. Here we take $\mathcal{C}=0.2$; for fibers of $r_{\rm c}=0.5$ (open diamonds), we find ${\rm {Re}}_{\rm {cr}}\approx 9.4$ and $\dot{\gamma} T_0\approx 17.9$; for fibers of $r_{\rm c}=0.1$ (filled circles), we find ${\rm {Re}}_{\rm {cr}}\approx 2.2$ and $\dot{\gamma} T_0\approx 30.6$. 
(c) Dependence of the tumbling period $T$ (normalized by the extrapolated period $T_0$ at $\mathcal{C}=0$) on $\mathcal{C}$. The period $T$ follows a power-law scaling (solid line): $\dot {\gamma} T \propto (\mathcal{C}_{\rm {cr}}-\mathcal{C})^{-1/2}$. Here we take ${\rm{Re}_{\rm ps}}=1.5$ and $r_{\rm c}=0.1$, and we find $\mathcal{C}_{\rm {cr}}\approx 0.7$ and $\dot{\gamma} T_0\approx 29.5$. The dimensions of the computational in the $x$ and $z$-directions are $H/d=50$ and $L_{\rm x}/d=25$, respectively.} }
\label{Fig:Rigid-Period}
\end{figure*}

\begin{figure*} [!htb]
\centering
\includegraphics[width= 2.0\columnwidth]{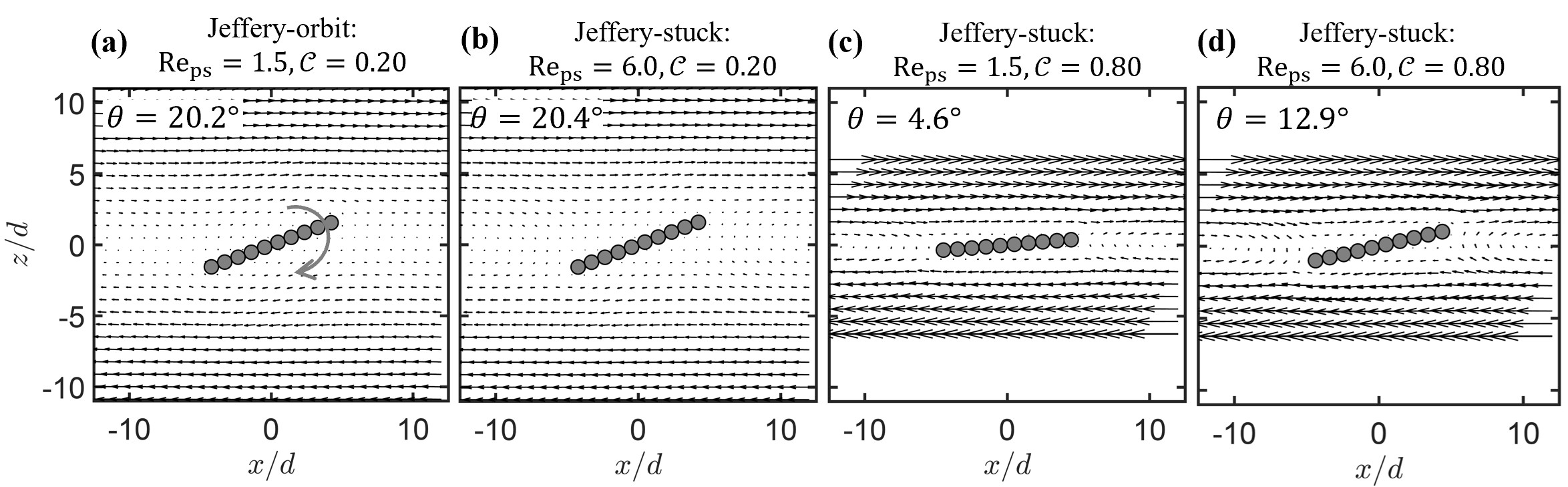}
\caption{\textcolor{black}{Flow fields around rigid fibers for different ${\rm{Re}_{ps}}$ and $\mathcal{C}$. (a) Jeffery-orbit for free rigid fibers under a small Reynolds number ${\rm{Re}_{ps}}=1.5$ and weak confinement $\mathcal{C}=0.2$. (b) Jeffery-stuck for rigid fibers: the fiber is stuck at an angle $\theta=20.4^{\rm o}$ under a large Reynolds number ${\rm{Re}_{ps}}=6.0$ and weak confinement $\mathcal{C}=0.2$. (c) Jeffery-stuck for rigid fibers: the fiber is stuck at an angle $\theta=4.6^{\rm o}$ under the small Reynolds number ${\rm{Re}_{ps}}=1.5$ and strong confinement $\mathcal{C}=0.8$.
(d) Jeffery-stuck for rigid fibers: the fiber is stuck at an angle $\theta=12.9^{\rm o}$ under the large Reynolds number ${\rm{Re}_{ps}}=6.0$ and strong confinement $\mathcal{C}=0.8$. 
In all cases, we take the flexible stiffness ${\cal K}=3.3 \times 10^0$ and the fiber aspect ratio to be $r_{\rm c}=1/N=0.1$ with $N$ being the number of particles in the fiber. The dimensions of the computational domain in the $x$ and $z$-directions are $H/d=N/\mathcal{C}$ and $L_{\rm x}/d=25$, respectively.}}
\label{Fig:Rigid-Flow}
\end{figure*}

We now study the dynamics of a single non-Brownian flexible fiber in 2D Couette flows. 
Consider a flexible fiber of contour length $\ell_{\rm c}$ immersed in a viscous \textcolor{black}{fluid} that is confined between two solid walls of distance $H$ as shown in Fig.~\ref{Fig:Method-Schematic}(b). \textcolor{black}{The horizontal $x$-dimension is denoted as $L_{\rm x}$.} Each flexible fiber is composed of $N$ identical rigid particles of radius $a$ and diameter $d=2a$, and hence $\ell_{\rm c}=Nd$. In this work, we only consider the highly symmetric case, in which the center of the fiber is placed on the center at $z=H/2$ between the two walls. The Couette flow is generated by the two walls moving at different velocities, $-V_0 \hat{\boldsymbol{x}}$ and $V_0 \hat{\boldsymbol{x}}$ at the bottom ($z=0$) and top ($z=H$) walls, respectively. No slip and impermeable conditions are employed on the surfaces of the two walls and the periodic condition is used in the lateral $x$-direction. 

The Couette flow or simple shear flow in $x-z$ plane with the shear rate $\dot{\gamma}=2V_0/H$ can be decomposed into two parts~\cite{du2019dynamics} (see Fig.~\ref{Fig:Method-Schematic}(c)): a rigid-body rotation with angular velocity $-\dot{\gamma}/2$ and a pure shear flow that consists of elongation along $\frac{\sqrt{2}}{2}(\hat{x}+\hat{z})$ direction and compression along $\frac{\sqrt{2}}{2}(-\hat{x}+\hat{z})$ direction, both of rate $\dot{\gamma}/2$. Therefore, the flexible fibers immersed in the symmetrical center of Couette flows will tumble (or rotate) and deform (or buckle), which are governed mainly by the following four dimensionless parameters 
\begin{equation}\label{Eq:Flexible-DimPara}
\mathcal{K}=\frac{2k_{\rm b}}{ \eta_{\rm f} \dot{\gamma} \ell_{\rm c}^{2}}, 
\quad 
\mathcal{C}=\frac{\ell_{\rm c}}{H}, \quad
{ \rm{Re}_{\rm {ps}}}= \frac{\rho  \dot{\gamma} \ell_{\rm c}^2}{ {4\eta_{\rm f}}}, \quad
r_{\rm c}=\frac{d}{\ell_{\rm c}}. 
\end{equation} 
These physical dimensionless parameters are related to and can be calculated from the dimensionless parameters used in our simulations as discussed in Sec.~\ref{Sec:Method-LBM}. Here we explain the physical meaning of each parameter as follows. 

(1) $0\leq \mathcal{K} < \infty$ measures the fiber stiffness to hydrodynamic shear force~\cite{nguyen2014hydrodynamics,slowicka2020flexible,yamamoto1993method,liu2018morphological,du2019dynamics}, which is the ratio of elastic restoring bending force (per length) $k_{\rm b}/\ell_{\rm c}$ to hydrodynamic shear force $\eta_{\rm f} \dot{\gamma} \ell_{\rm c}/2$. Alternatively, $\mathcal{K}$ can be regarded as the ratio of the characteristic Couette flow time, $\tau_{\rm f}=(\dot{\gamma}/2)^{-1}$, to the elastic relaxation time of a bending deformation, $\tau_{\rm b}={\eta_{\rm f} \ell_{\rm c}^2}/{k_{\rm b}}$. Particularly, $\mathcal{K} \to \infty$ and $\mathcal{K} \to 0$ correspond to the limits of rigid fibers and freely-jointed fibers without bending stiffness, respectively.

(2) $0 \leq \mathcal{C} \leq 1 $ measures the strength of the confinement, which is the ratio of the fiber contour length $\ell_{\rm c}$ to the distance $H$ between the two walls. Particularly, $\mathcal{C} \to 0$ corresponds to the limit of free (no confinement) fiber dynamics in shear flows. $\mathcal{C} \to 1$ or $\ell_{\rm c}\to H$ corresponds to the strongest confinement case considered in this work. 

(3) ${\rm{Re}_{\rm {ps}}}$ is the characteristic particle Reynolds number in shear flows~\cite{zettner2001moderate,huang2012rotation}, which is the ratio of inertial forces $\rho (\dot{\gamma} \ell_{\rm c}/2)^2 (\ell_{\rm c}/2)$ to hydrodynamic shear force $\eta_{\rm f} \dot{\gamma} \ell_{\rm c}/2$. 

(4) $r_{\rm c}$ is the (width-to-length) aspect ratio. Since $\ell_{\rm c}=Nd$, we have $r_{\rm c}=1/N$. For example, as shown in Fig.~\ref{Fig:Rigid-Flow}, the aspect ratio of the fiber composed of $N=10$ beads is $r_{\rm c}=0.1$. 

\subsection{Dynamics of rigid fibers: effects of confinement strength and Reynolds number}\label{Sec:Flexible-Rigid} 


We first consider the dynamics of rigid fibers (in the limit of $\mathcal{K}\to \infty$) in shear (Couette) flows as shown schematically in Fig.~\ref{Fig:Method-Schematic}(b), which have been extensively studied since G.B. Jeffery in 1922~\cite{jeffery1922motion}. Most previous works as Jeffery did study the dynamics of rigid fibers or anisotropic particles under the limits of zero Reynolds number ${\rm Re_{ps}} \to 0$ and usually neglected the effects of confinement, \emph{i.e.}, ${\mathcal C}\to 0$. In contrast, here our focus will be placed on the less-studied effects of the confinement strength (for $0< {\mathcal C} < 1$) and the finite Reynolds number (for $1<{\rm Re_{ps}}< 10$).

In the limits of ${\rm Re_{ps}} \to 0$ and ${\mathcal C}\to 0$, Jeffery predicted~\cite{jeffery1922motion} that anisotropic (elliptical) particles tumble in simple shear flows and the angular velocity $\omega$ follows the formula of now so-called ``Jeffery orbit"~\cite{jeffery1922motion,zhang2015anisotropic}:
\begin{equation}\label{Eq:Flexible-omega}
\omega= \frac{\dot{\gamma}}{2} \left[ \frac{1-r_{\rm c}^2}{1+r_{\rm c}^2} \cos (2\theta) - 1 \right]
\end{equation}
with $r_{\rm c}=d/\ell_{\rm c}$ being the aspect ratio of anisotropic fiber (or particle), defined in Eq.~(\ref{Eq:Flexible-DimPara}).
In our simulations, we take a large enough $\mathcal{K}\gg 1$ such that the fibers behave like rigid fibers and do not bend in shear flows. As shown in Fig.~\ref{Fig:Rigid-Period}(a), the angular velocity $\omega$ of the tumbling rigid fiber follows Jeffery orbit. A least-square fitting using Eq.~(\ref{Eq:Flexible-omega}) gives an ``effective aspect ratio" $r_{\rm c}\approx 0.132$, which is very close to the ``geometrical aspect ratio" $r_{\rm c}=0.1$ of the fiber. The small difference can be attributed to the difference between the tumbling of rigid rod-like fibers and that of elliptical particles that Jeffery considered in his original work~\cite{jeffery1922motion}. 

\textcolor{black}{One general characteristic parameter of the tumbling dynamics of flexible fibers is the tumbling period $T$, which can be defined as the time interval that the fiber rotates an angle of $180^{\rm o}$ (not $360^{\rm o}$ because the period of $\cos (2\theta)$ in $\omega$ is $180^{\rm o}$). Then from $\omega$ of Jeffery orbit (with ${\rm Re_{ps}} \to 0$ and ${\mathcal C}\to 0$) in Eq.~(\ref{Eq:Flexible-omega}), we can easily see that $T$ is only a function of shear rate $\dot{\gamma}$ and aspect ratio $r_{\rm c}$. However, when we increase ${\rm Re_{ps}}$ and ${\mathcal C}$, we find $T$ shows strong nonlinear dependence on ${\rm Re_{ps}}$ and ${\mathcal C}$.} 
In Fig.~\ref{Fig:Rigid-Period}(b), we show that $T$ of the rigid fibers diverges as ${\rm Re_{ps}}$ increases to some critical value ${\rm {Re}}_{\rm {cr}}$. That is, when ${\rm Re_{ps}} > {\rm {Re}_{\rm {cr}}}$, the rigid fiber gets stuck in some direction and becomes stationary in the steady-state Couette flow. For ${\rm Re_{ps}}<{\rm {Re}_{\rm {cr}}}$, $T$ diverges with increasing ${\rm Re_{ps}}$, following the universal scaling law: $\dot {\gamma} T \propto ({\rm {Re}}_{\rm {cr}} - {\rm Re_{ps}})^{-1/2}$, which has been reported before in both experiments and simulations~\cite{ding2000dynamics}. Moreover, we find ${\rm {Re}}_{\rm {cr}}$ increases with the fiber aspect ratio $r_{\rm c}=1/N$ as observed by Zettner and Yoda~\cite{zettner2001moderate}. 
In addition, interestingly, we note in Fig.~\ref{Fig:Rigid-Period}(c) that when the confinement is weak, $\mathcal{C}<0.5$, the tumbling period $T$ shows a very weak dependence on $\mathcal{C}$. However, when $\mathcal{C}\geq 0.5$, $T$ shows a similar divergence dependence on $\mathcal{C}$ as on  ${\rm Re_{ps}}$: $T$ diverges as $\mathcal{C}$ increases to some critical value ${\mathcal C}_{\rm {cr}}$ (around $0.71$), following the power-law scaling: $\dot {\gamma} T \propto (\mathcal{C} _{\rm {cr}} - \mathcal{C})^{-1/2}$. When $\mathcal{C} > \mathcal{C} _{\rm {cr}}$, the rigid fiber also gets stuck in some direction and becomes stationary in the steady-state Couette flow. 

To better understand how the rigid fiber gets stuck, we visualize the flow fields around the rigid fibers (in Fig.~\ref{Fig:Rigid-Flow}) under different ${\rm Re_{ps}}$ and $\mathcal{C}$. The prescribed background simple shear flow is perturbed mostly near the rigid fibers and two regions can be identified: the simple shear region near the walls and far away from the fiber; the recirculation region at the center around the fiber. Similar patterns have been obtained in the previous experiments~\cite{zettner2001circular} and simulations~\cite{ding2000dynamics}. The shear region contributes a positive torque on the fiber, driving the clockwise fiber rotation (as schematically shown in Fig.~\ref{Fig:Method-Schematic}(c)), while the recirculating region has a negative (counter-clockwise) contribution~\cite{ding2000dynamics}, resisting the clockwise rotation of the fiber. 
\textcolor{black}{By comparing the flow fields under different ${\rm Re_{ps}}$ and $\mathcal{C}$, we can easily understand the mechanism underlying the stuck of rigid fibers at large ${\rm Re_{ps}}$ and/or strong $\mathcal{C}$. When ${\rm Re_{ps}}$ and/or $\mathcal{C}$ increases, the size of the central recirculation region increases, and their contribution to the counter-clockwise torque becomes larger, which consequently results in the increase of the clockwise rotation period of the rigid fiber, as shown in Fig.~\ref{Fig:Rigid-Period}(b,c). Particularly, when ${\rm Re_{ps}}>{\rm Re_{\rm cr}}$ and/or $\mathcal{C}>\mathcal{C}_{\rm cr}$, the counter-clockwise torques resulted from the central recirculating flows balance the clockwise rotating torques applied by the simple shear flows near the wall, as shown in Fig.~\ref{Fig:Rigid-Flow}(b,c,d). The fiber gets stuck for large enough ${\rm Re_{ps}}$, or $\mathcal{C}$, or both, however, the stuck angle shows a non-monotonic dependence on the magnitude of ${\rm Re_{ps}}$ and $\mathcal{C}$. }

\subsection{Dynamics of flexible fibers: effects of fiber stiffness, confinement strength, and Reynolds number}\label{Sec:Flexible-Dynamics}

\begin{figure*} [!htb]
\centering
\includegraphics[width=1.8\columnwidth]{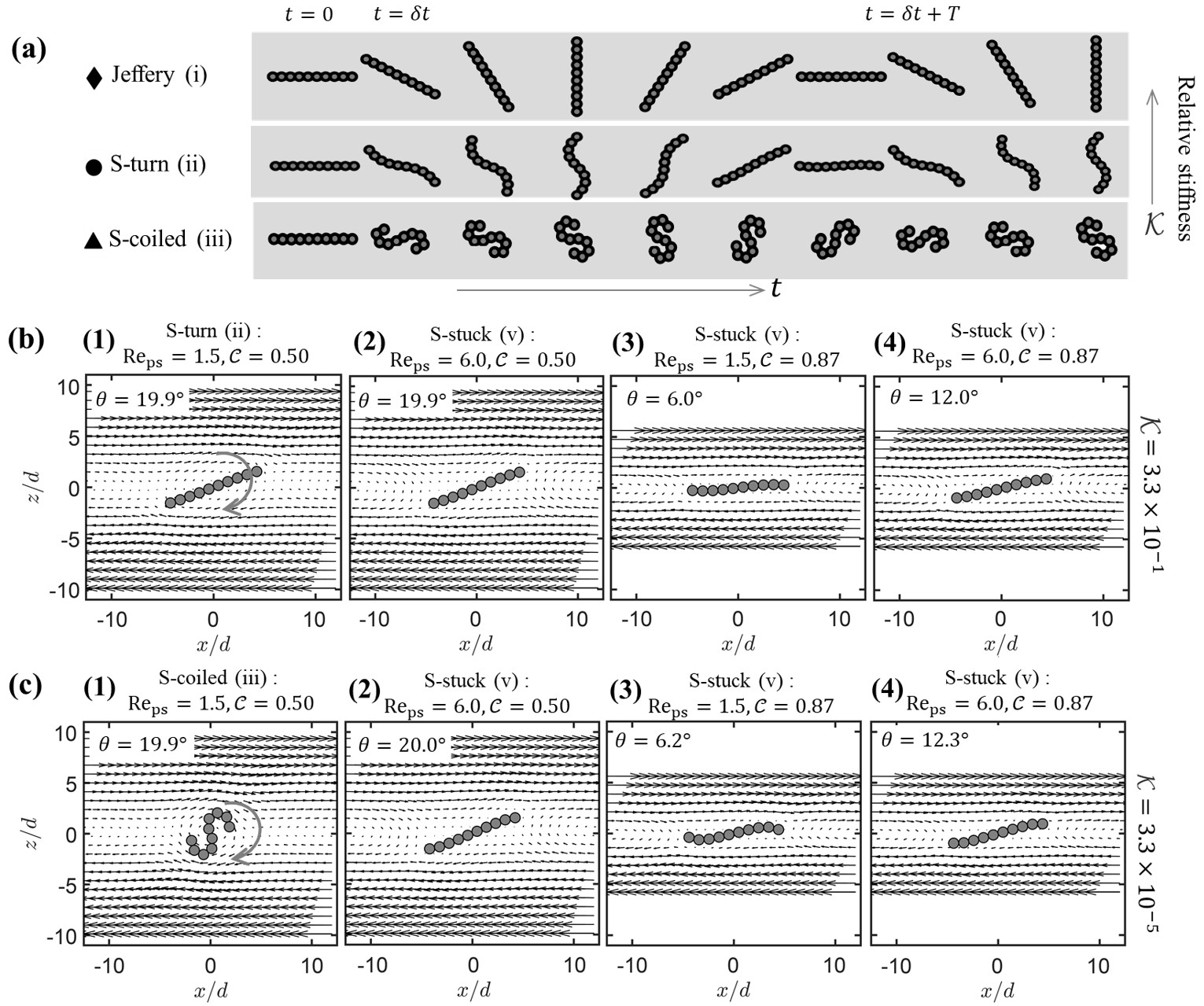}
\caption{\textcolor{black}{Dynamics of flexible fibers of different stiffnesses $\mathcal{K}$ in confined shear (Couette) flows. (a) Three different tumbling orbits were identified according to the fiber shape dynamics: (i) Jeffery orbit of rigid fibers with $\mathcal{K}=3.3 \times \textcolor{black}{ 10^0}$, (ii) S-turn orbit of slightly flexible fibers with $\mathcal{K}=3.3 \times \textcolor{black}{10^{-1}}$, and (iii) S-coiled orbit of fairly flexible fibers with $\mathcal{K}=3.3 \times 10^{-5}$. 
(b,c) Flow fields around flexible fibers for different ${\rm{Re}_{ps}}$ and $\mathcal{C}$. 
(b-1) S-turn orbit of slightly flexible fibers under a small Reynolds number ${\rm{Re}_{ps}}=1.5$ and weak confinement $\mathcal{C}=0.5$. (b-2) S-stuck (regime v): the fiber is stuck at an angle $\theta=19.9^{\rm o}$ under a large Reynolds number ${\rm{Re}_{ps}}=6.0$ and weak confinement $\mathcal{C}=0.5$. (b-3) S-stuck: the fiber is stuck at an angle $\theta=6.0^{\rm o}$ under the small Reynolds number ${\rm{Re}_{ps}}=1.5$ and strong confinement $\mathcal{C}=0.87$.
(b-4) Jeffery-stuck for rigid fibers: the fiber is stuck at an angle $\theta=12.0^{\rm o}$ under the large Reynolds number ${\rm{Re}_{ps}}=6.0$ and strong confinement $\mathcal{C}=0.87$. 
(c-1) S-coiled orbit for fairly flexible fibers under a small Reynolds number ${\rm{Re}_{ps}}=1.5$ and weak confinement $\mathcal{C}=0.5$. 
(c-2) S-stuck (regime v): the fiber is stuck at an angle $\theta=20.0^{\rm o}$ under a large Reynolds number ${\rm{Re}_{ps}}=6.0$ and weak confinement $\mathcal{C}=0.5$. 
(c-3) S-stuck (regime v): the fiber is stuck at an angle $\theta=6.2^{\rm o}$ under the small Reynolds number ${\rm{Re}_{ps}}=1.5$ and strong confinement $\mathcal{C}=0.87$.
(c-4) S-stuck (regime v): the fiber is stuck at an angle $\theta=12.3^{\rm o}$ under the large Reynolds number ${\rm{Re}_{ps}}=6.0$ and strong confinement $\mathcal{C}=0.87$. 
In all cases, we take the fiber aspect ratio to be $r_{\rm c}=1/N=0.1$ with $N$ being the number of particles in the fiber. The dimensions of the computational domain in the $x$ and $z$-directions are $H/d=N/\mathcal{C}$ and $L_{\rm x}/d=25$, respectively.}
}\label{Fig:Flexible-Shape}
\end{figure*}

\begin{figure*} [!htb]
\centering
\includegraphics[width=1.670\columnwidth]{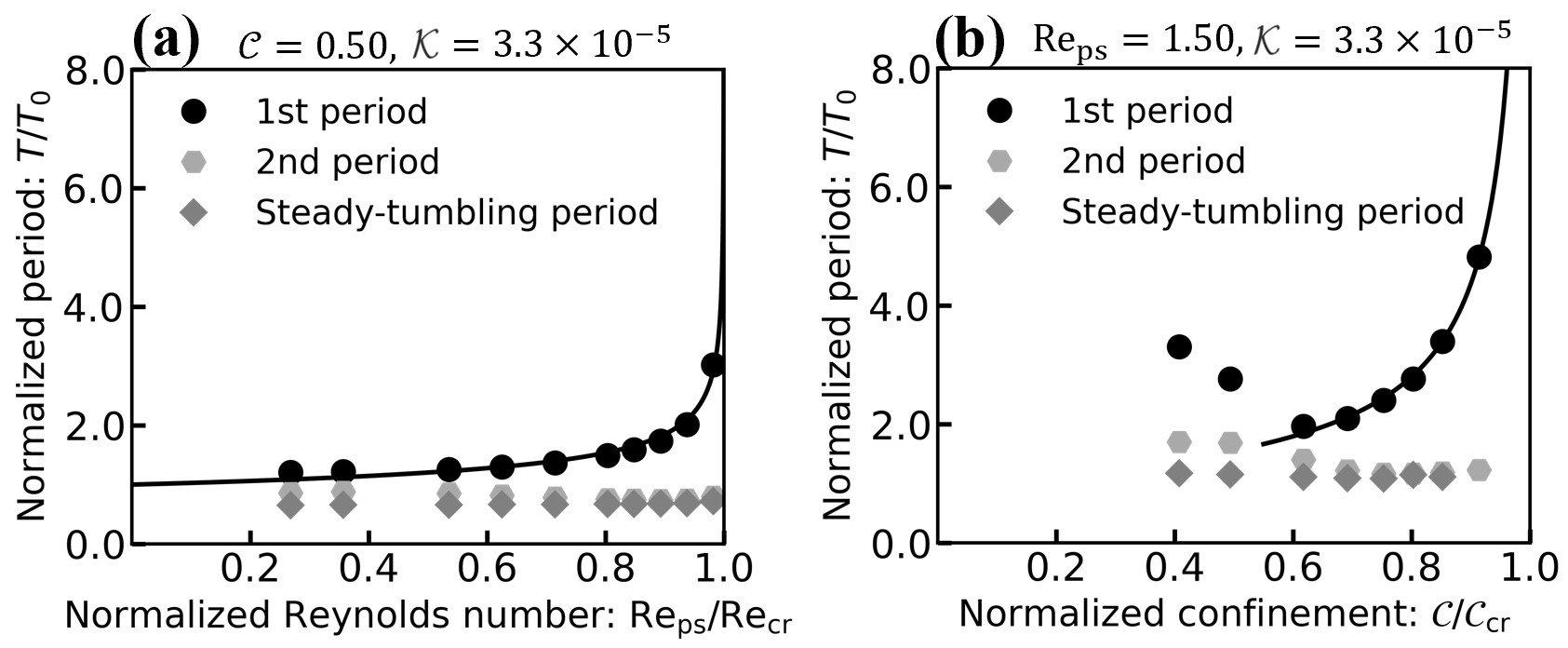}
\caption{\textcolor{black}{Tumbling dynamics of fairly flexible fibers (the S-coiled orbit of fibers with $\mathcal{K}=3.3 \times 10^{-5}$).
(a) Dependence of the tumbling period $T$ (normalized by the extrapolated period $T_0$ at ${\rm{Re}_{ps}}=0$) on ${\rm{Re}_{ps}}$. 
The period $T$ follows a power-law scaling (solid line): $\dot {\gamma} T \propto ({\rm {Re}}_{\rm {cr}} -{\rm{Re}_{\rm ps}})^{-0.3}$. Here we take $\mathcal{C}=0.5$, $r_{\rm c}=0.1$, and we find ${\rm {Re}}_{\rm {cr}}\approx 5.6$ and $\dot{\gamma} T_0\approx 11.9$. 
(b) Dependence of the tumbling period $T$ (normalized by the extrapolated period $T_0$ at $\mathcal{C}=0$) on $\mathcal{C}$. The period $T$ follows a power-law scaling (solid line): $\dot {\gamma} T \propto (\mathcal{C}_{\rm {cr}}-\mathcal{C})^{-0.6}$. Here we take ${\rm{Re}_{\rm ps}}=1.5$ and $r_{\rm c}=0.1$, and we find $\mathcal{C}_{\rm {cr}}\approx 0.8$ and $\dot{\gamma} T_0\approx 7.3$.
} }
\label{Fig:Flexible-Period}
\end{figure*}

\begin{figure*} [!htb]
\centering
\includegraphics[width= 2.0\columnwidth]{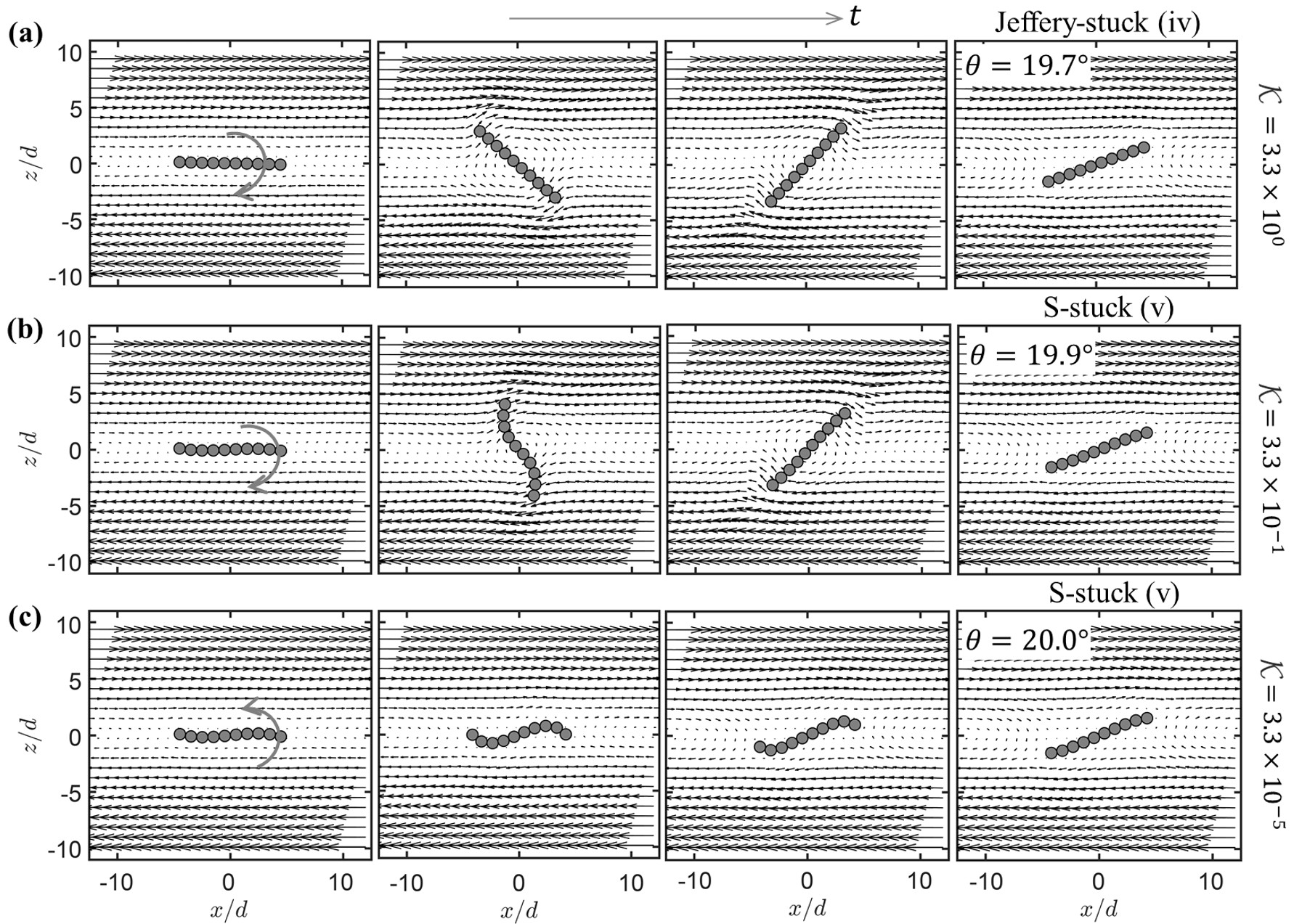}
\caption{\textcolor{black}{Stuck pathway of fibers of different stiffnesses under a large Reynolds number ${\rm{Re}_{ps}}=6.0$ and weak confinement $\mathcal{C}=0.5$, starting from the same initial flat horizontal orientation ($t=0$ as shown in Fig.~\ref{Fig:Flexible-Shape}(a)). The rigid fibers in (a) and the slightly flexible fibers in (b) both rotate clockwise for about $160^{\rm o}$ before being stuck at an angle around $\theta \approx 20^{\rm o}$. However, the fairly flexible fiber in (c) rotates counter-clockwise for about $20^{\rm o}$ and gets stuck there. The dimensions of the computational domain in the $x$ and $z$-directions are $H/d=N/\mathcal{C}$ and $L_{\rm x}/d=25$, respectively.}}
\label{Fig:Stuck-Pathway}
\end{figure*}

\begin{figure*} [!htb]
\centering
\includegraphics[width=1.75\columnwidth]{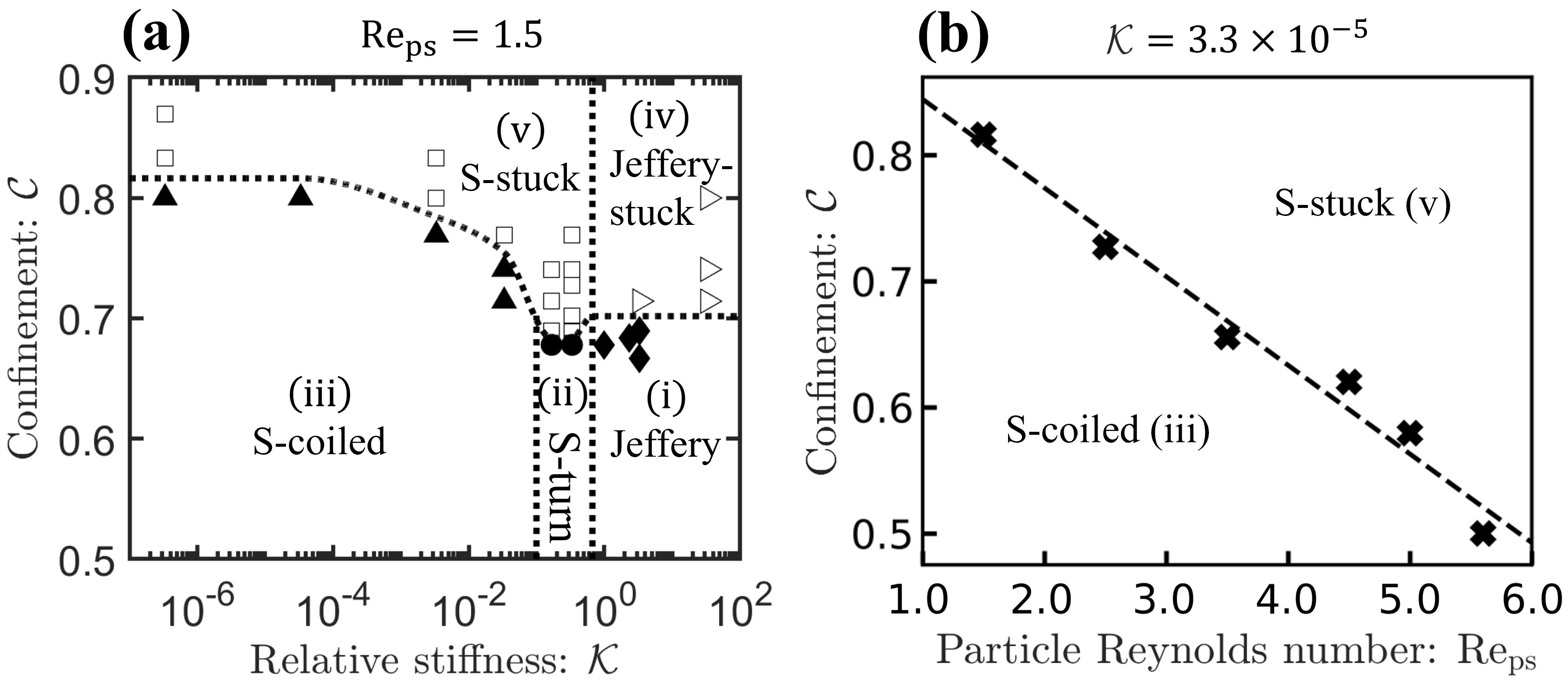}
\caption{\textcolor{black}{(a) Shape diagram of flexible fibers as a function of fiber stiffness $\mathcal{K}$ and confinement strength $\mathcal{C}$ at a given Reynolds number ${\rm Re_{ps}}=1.5$. For small $\mathcal{C}$ or weak confinement, three different types of dynamic orbits have been identified as shown in Fig.~\ref{Fig:Flexible-Shape}. (i) Jeffery orbits of rigid fibers for very large $\mathcal{K}$. (ii) S-turn orbit of slightly flexible fibers for intermediate $\mathcal{K}$ where the fiber tumbling shows a periodic shape-changing dynamic pattern. (iii) S-coiled orbits of fairly flexible fibers for small $\mathcal{K}$ where the fiber folds to an S-shape and tumbles periodically and steadily. If $\mathcal{C}$ increases over some critical value $\mathcal{C}_{\rm cr}$ (dotted lines), the fiber gets stuck along some direction for rigid fibers in (iv) Jeffery-stuck regime and for soft fibers in (v) S-Stuck regime, respectively.  
(b) Shape diagram of fairly flexible fibers as a function of the particle Reynolds number ${\rm Re_{ps}}$ and the confinement $\mathcal{C}$ at a given fiber stiffness $\mathcal{K} = 3.3 \times 10^{-5}$. The regime interface is denoted by cross markers and the fitted dashed line. } }
\label{Fig:Flexible-Shape2}
\end{figure*}

Rigid fibers have been shown to tumble or rotate in Couette flows without deformation and the tumbling period $T$ is found to show a strong dependence on the Reynolds number $\rm Re_{ps}$ and confinement strength $\mathcal{C}$. Here we consider the dynamics of flexible fibers in Couette flows. In contrast to the dynamics of rigid fibers, flexible fibers would not only tumble but can also deform significantly. Our focus will be placed on the flow-induced shape changes and the tumbling dynamics of flexible fibers as well as their dependence on the fiber stiffness $\mathcal{K}$, the confinement strength $\mathcal{C}$, and the Reynolds number $\rm{Re}_{ps}$. 


We first carried out a systematic study about the effects of fiber stiffness $\mathcal{K}$ on highly symmetrical flexible fibers under weak confinement with small ${\mathcal C}$ at a small Reynolds number ${\rm Re_{ps}}=1.5$. 
Three major types of tumbling orbits can be identified according to different fiber stiffnesses $\mathcal{K}$ for small ${\mathcal C}$ as shown in the snapshots and flow fields of our simulations in Fig.~\ref{Fig:Flexible-Shape} and in the shape diagram summarized in Fig.~\ref{Fig:Flexible-Shape2}. 
\begin{itemize}
\item[\textcolor{black}{(i)}:] {\emph{Jeffery orbits of rigid fibers}.} When $\mathcal{K}$ is very large $\gg 1$, the fiber behaves like rigid bodies and tumbles periodically (following Jeffery orbits) without any visible deformation as analyzed in Sec.~\ref{Sec:Flexible-Rigid}. \textcolor{black}{The flow field around the tumbling rigid fiber is shown in Fig.~\ref{Fig:Rigid-Flow}(a).}   

\item[\textcolor{black}{(ii)}:] {\emph{S-turn orbits of slightly flexible fibers}.} \textcolor{black}{When $\mathcal{K}$ is smaller than some threshold value (of the order of magnitude of $1$ according to simple scaling analysis: when the hydrodynamic shear force $\eta_{\rm f} \dot{\gamma} \ell_{\rm c}/2$ is comparable to the elastic restoring bending force $k_{\rm b}/\ell_{\rm c}$), the fiber is slightly flexible and its both ends are bent simultaneously in opposite directions so that the fiber is bent to an S-shape, as schematically explained in Fig.~\ref{Fig:Method-Schematic}(c)). The end-to-end distance of the fiber decreases sharply from the fiber contour length $\ell_{\rm c}$ (undeformed rigid fibers) to $\sim 10\%$ smaller than $\ell_{\rm c}$. However, when the fiber orients to the angle around $\theta=45^{\rm o}$ (which is the direction of the maximal principal tensile stress), the fiber is straightened again as shown in Fig.~\ref{Fig:Flexible-Shape}.} Such S-shape fiber bending dynamics have been predicted by Forgacs and Mason~\cite{forgacs1959particle} for highly symmetrical fibers as in our case, and according to their classification of fiber bending shapes, this type of orbit can be termed as "S-turn orbit". 
\textcolor{black}{The flow field around the tumbling slightly flexible fiber is shown in Fig.~\ref{Fig:Flexible-Shape}(b-1). }  

\item[\textcolor{black}{(iii)}:] {\emph{S-coiled orbits of fairly flexible fibers}.} For fairly flexible fibers of very small $\mathcal{K}$ ($\ll 1$), the fiber is folded to an S-shape quickly and tumbles periodically and steadily. In contrast to the S-turn orbit, the fairly flexible fiber is no longer straightened during its tumbling but is kept to be at the S-coiled configuration. This type of orbit can be termed an "S-coiled orbit". \textcolor{black}{The flow field around the tumbling fairly flexible fiber is shown in Fig.~\ref{Fig:Flexible-Shape}(c-1). }  
\end{itemize}

\textcolor{black}{We then quantify the effects of particle Reynolds number ${\rm Re_{ps}}$ and confinement strength $\mathcal{C}$ on the tumbling dynamics of flexible fibers by studying the dependence of the tumbling period $T$ on ${\rm Re_{ps}}$ and ${\mathcal C}$. As shown in Fig.~\ref{Fig:Flexible-Period}, we find that $T$ measured in the first tumbling period of fairly flexible fibers shows very similar diverging behaviors to those of rigid fibers (shown in Fig.~\ref{Fig:Rigid-Period}). Fig.~\ref{Fig:Flexible-Period}(a) shows that as ${\rm Re_{ps}}$ increases to some critical value ${\rm {Re}}_{\rm {cr}}$ (where the fiber gets stuck), $T$ diverges with increasing ${\rm Re_{ps}}$, also following a weaker diverging power-law scaling: $\dot {\gamma} T \propto ({\rm {Re}}_{\rm {cr}} - {\rm Re_{ps}})^{-0.3}$. Fig.~\ref{Fig:Flexible-Period}(b) shows that $T$ diverges as $\mathcal{C}$ increases to some critical value ${\mathcal C}_{\rm {cr}}$ (where the fiber gets stuck), following another stronger diverging power-law scaling: $\dot {\gamma} T \propto (\mathcal{C} _{\rm {cr}}-\mathcal{C})^{-0.6}$.
However, in the second and afterward tumbling periods, the fairly flexible fibers fold to an S-coiled shape and $T$ becomes almost a constant independent on both ${\rm Re_{ps}}$ and ${\mathcal C}$. This can be understood if we think the S-coiled fibers behave effectively as a much shorter fiber (with a smaller length $\ell_{\rm eff}$ and a larger aspect ratio) and hence the effective Reynolds number ${\rm Re_{\rm eff}}= {\rho  \dot{\gamma} \ell_{\rm eff}^2}/{ {4\eta_{\rm f}}}$ and effective confinement strength ${\mathcal C}_{\rm eff}={\ell_{\rm eff}}/{H}$ are both much smaller than those defined and set in terms of fiber contour length $\ell_{\rm c}$. } 

\textcolor{black}{Furthermore, to understand how flexible fibers get stuck at large ${\rm Re_{ps}}$ and $\mathcal{C}$, we have visualized and compared the flow fields around the flexible fibers (in Fig.~\ref{Fig:Flexible-Shape}) under different ${\rm Re_{ps}}$ and $\mathcal{C}$. We find that the pattern of the flow fields is very similar to those of rigid fibers as shown in Fig.~\ref{Fig:Rigid-Flow} and discussed in Sec.~\ref{Sec:Flexible-Rigid}. Therefore, we can understand the stuck of flexible fibers in the same manner as that of rigid fibers. When ${\rm Re_{ps}}>{\rm Re_{\rm cr}}$ and/or $\mathcal{C}>\mathcal{C}_{\rm cr}$, the counter-clockwise torques resulted from the central recirculating flows balance the clockwise rotating torques applied by the simple shear flow near the wall, as shown in Fig.~\ref{Fig:Flexible-Shape}(b,c,d).}

\textcolor{black}{In addition, we have also checked how fibers of different stiffnesses at large $\rm{Re}_{ps}$ get stuck when starting from the same initial flat horizontal orientation (at $t=0$ as shown in Fig.~\ref{Fig:Flexible-Shape}(a)). As shown in Fig.~\ref{Fig:Stuck-Pathway}(a,b), we find that for rigid and slightly flexible fibers, the fibers both rotate clockwise for about $160^{\rm o}$ before being stuck at an angle around $\theta \approx 20^{\rm o}$. In contrast, fairly flexible fibers rotate counter-clockwise for about $20^{\rm o}$ and get stuck there as shown in Fig.~\ref{Fig:Stuck-Pathway}(c). Such peculiar counter-clockwise rotation can be understood simply if we track the initial fiber shape changes and the flow fields surrounding the fiber as shown in the first two snapshots of Fig.~\ref{Fig:Stuck-Pathway}(c). For fairly flexible fibers at large $\rm{Re}_{ps}$, the counter-clockwise torque applied by the central recirculating flows is large enough to induce significant fiber buckling (of cosine S-shape) at the very beginning of their horizontal orientation (see the second snapshot of Fig.~\ref{Fig:Stuck-Pathway}(c)). The counter-clockwise torque applied on such cosine S-shape fibers by central recirculating flows dominates over the clockwise torque applied by the upper/bottom simple shear flows, thus leading to the peculiar counter-clockwise pathway toward the stuck steady orientation. On the other hand, for fibers of any stiffnesses under strong confinement (either small or large $\rm{Re}_{ps}$), all fibers are found to rotate counter-clockwise to their final stuck orientation. That is, under strong confinement, the counter-clockwise torque applied by the central recirculating flows always dominates over the clockwise torque applied by the upper/bottom simple shear flows.}


\textcolor{black}{Finally, the different steady dynamic regimes discussed above are summarized in shape diagrams according to the three major dimensionless parameters, the fiber stiffness $\mathcal{K}$, the confinement strength $\mathcal{C}$, and the Reynolds number ${\rm{Re}_{\rm {ps}}}$. In Fig.~\ref{Fig:Flexible-Shape2}(a), a shape diagram is obtained for a large range of $\mathcal{K}$ and $\mathcal{C}$ at a small $\rm{Re}_{ps}=1.5$. For fibers under weak confinement, three distinct tumbling orbits have been identified according to $\mathcal{K}$. As $\mathcal{C}$ increases over some $\mathcal{K}$-dependent critical value $\mathcal{C}_{\rm cr}$, two fiber-stuck (Jeffery-stuck and S-stuck) regimes are then identified. 
In Fig.~\ref{Fig:Flexible-Shape2}(b), we consider the fairly flexible fibers (with a given small $\mathcal{K}=3.3\times 10^{-5}$) and obtain a shape diagram for a large range of $ {\rm{Re}_{\rm {ps}}}$ and $\mathcal{C}$. The dashed line separating the S-coiled and S-stuck regimes can be understood from the following two different perspectives. (1) The critical confinement $\mathcal{C}_{\rm cr}$ over which flexible fibers get stuck decreases with increasing $ {\rm{Re}_{\rm {ps}}}$. (2) The critical particle Reynolds number $\rm{Re}_{\rm cr}$ over which flexible fibers get stuck decreases with increasing $\mathcal{C}$. That is, the fiber's periodic tumbling is hindered by increasing the particle Reynolds number ${\rm{Re}_{\rm {ps}}}$ or the confinement strength $\mathcal{C}$, or both. }

\section{Concluding remarks}\label{Sec:Conclusions}

In this work, the fluid particle dynamics (FPD) method originally developed for rigid particles in viscous \textcolor{black}{fluids} is extended using the bead-spring model of flexible fibers to study the dynamics of non-Brownian fibers in 2D confined shear flows (\emph{i.e}, Couette flows). The FPD method is implemented by a multiple–relaxation–time (MRT) scheme of the lattice Boltzmann method (LBM). The numerical scheme is validated firstly by two sets of benchmark simulations that involve \textcolor{black}{fluid}-solid coupling: (1) steady-state velocity fields of Couette flows or simple shear flows between two solid walls moving at different velocities, and (2) the drag forces applied on the circular particle moving in viscous \textcolor{black}{fluids} for various Reynolds numbers ranging from $1$ to $10$. Practically, we find from these benchmark simulations that the computational accuracy of our numerical simulations is high enough if $r_\eta=\eta_{\rm s}/\eta_{\rm f} \ge 50$ and $r_\xi=\xi/d \le 0.125$. To ensure accuracy and save computational cost, we then set $r_\eta=100$ and $r_\xi=0.125$ in most of the simulations for the dynamics of flexible fibers in Couette flows confined between two solid walls of distance $H$. 

Physically, our focus is placed on the effects of the fiber stiffness $\mathcal{K}$, the confinement strength $\mathcal{C}$, and the Reynolds number $ {\rm{Re}_{\rm {ps}}}$. A shape diagram for the fiber moving in 2D confined shear flows is obtained in Fig.~\ref{Fig:Flexible-Shape2} for a large range of fiber stiffness $\mathcal{K}$ and confinement strength $\mathcal{C}$. For fibers under weak confinement, three distinct orbits have been identified according to $\mathcal{K}$. (1) Jeffery orbits of rigid fibers (large $\mathcal{K}$). The fibers behave like rigid bodies and tumble periodically without any visible deformation as analyzed in Sec.~\ref{Sec:Flexible-Rigid}. (2) S-turn orbits of slightly flexible fibers (intermediate $\mathcal{K}$). The fiber is bent simultaneously to an S-shape and is straightened again \textcolor{black} {when it orients to an angle of around $\theta=45^{\rm o}$ relative to the $+\hat{\bf x}$-direction}. (3) S-coiled orbits of fairly flexible fibers (small $\mathcal{K}$). The fiber is folded to an S-shape and tumbles periodically and steadily. During the rotation, the fiber is no longer straightened but kept to be at the S-coiled configuration. In addition, we find that the tumbling period $T$ of all fibers shows similar dependence on the confinement strength $\mathcal{C}$ as shown in Fig \ref{Fig:Rigid-Period}(c) and discussed in Sec.~\ref{Sec:Flexible-Rigid}. When $\mathcal{C}$ is small, the period $T$ shows a very weak dependence on $\mathcal{C}$. When $\mathcal{C}$ increases over some critical value $\mathcal{C} _{\rm {cr}}$, the fibers are found to be stuck in some direction and become stationary in the steady-state Couette flow.

Finally, we make some general remarks as follows. 

\textcolor{black}{(1) We have used the LBM scheme to implement the extended FPD method for fiber dynamics. The LBM scheme based on microscopic models and mesoscopic kinetic equations has many advantages such as clear physical pictures, easy implementation, and fully parallel algorithms. However, LBM is not suitable for dynamics at small Reynolds numbers, the increase in computational accuracy requires lots of additional effort, and the large viscosity ratios and viscosity gradients present in the FPD method pose great challenges in the computational efficiency in the LBM. For example, in the simulations presented in Fig.~\ref{Fig:Flexible-Shape}, our numerical code based on LBM is very slow: it took about $24.5$ hours, $27.5$ hours, and $19.5$ hours to simulate one tumbling period of Jeffery orbits, S-turn orbits, and S-coiled orbits, respectively. Therefore, we are now trying other traditional numerical methods such as finite difference and finite element methods to implement the FPD method at low Reynolds numbers, in dense fiber suspensions, and in cases where the requirements for computational accuracy are very high.} 

(2) \textcolor{black}{We have only considered the dynamics of inextensible flexible fibers. However, the extensibility may have important effects on the buckling dynamics of flexible fibers in viscous flows. This can be considered easily using the extended FPD method in the future by considering small compress stiffness parameter, ${\mathcal K}_{\rm s}$ and introducing overlapping between neighbor particles in the fibers so that the fibers are extensible but still no gaps form between neighbor particles to avoid fluid penetration across the fibers.} Furthermore, the excluded volume interactions between fiber particles that are not connected by springs have not been considered. However, particle overlapping between particles has not been observed in all our simulations due to potentially strong repulsive hydrodynamic interactions between particles that are very close to each other. 

(3) We have only considered the highly symmetric fiber dynamics, in which case the fibers are placed on the symmetry center of Couette flows. Our simulations are carried out for single short fibers in 2D. The extensions to fibers off-center, to long fibers, to multiple fibers, and to 3D will be interesting. Fibers off center and long flexible fibers are known to show more deformation patterns and dynamic orbits~\cite{liu2018morphological} such as springy turn~\cite{forgacs1959particle,feng1994direct}, snake-turn orbits~\cite{forgacs1959particle,feng1994direct}, coil-stretch transitions~\cite{kantsler2012fluctuations,young2007stretch}, and knotting~\cite{kuei2015dynamics}. For multiple fibers, the many-body hydrodynamic interactions between flexible fibers are relevant to more dense fiber suspensions which will exhibit rich collective dynamics. Moreover, in 3D, the fiber will not only bend but also twist and the torsion becomes important. 

(4) Our numerical method can be further extended to study the dynamics of Brownian fibers in viscous flows and fibers in complex fluids such as multiphase flows and fluids with an internal degree of freedom~\cite{tanaka2018physical}. \textcolor{black}{Moreover, we can also combine the FPD method with coarse-grained membrane models consisting of many interacting particles to study the dynamics of ring-like polymers, capsules, and red blood cells in viscous flows~\cite{schmidt2022capsules}.}

    
    





\appendix

\section*{Acknowledgements}
Q. He is supported partly by the National Natural Science Foundation of China (No.11971020). 
X. Xu is supported partly by the Provincial Science Foundation of Guangdong (2019A1515110809), the National Natural Science Foundation of China (NSFC, No.~12131010), and the Guangdong Basic and Applied Basic Research Foundation (2020B1515310005). \textcolor{black}{ The work was carried out at the Shanxi Supercomputing center of China, and the calculations were performed on TianHe-2.}

\section*{References}
\bibliography{references}
\end{document}